\documentclass{jfm2}

\usepackage{hyperref}
\usepackage{graphicx}
\usepackage{natbib}
\usepackage{amsmath}
\usepackage{amssymb}

\usepackage{amsbsy}

\newcommand{\vect}[1]{\boldsymbol #1}
\newcommand{\boldnabla}{\boldsymbol\nabla}
\newcommand{\tensor}[1]{\mathsfbi{#1}}

\usepackage{upmath}

\title[The Surface Topography of a Magnetic Fluid]{The Surface Topography of a Magnetic Fluid
\textemdash\ \\ 
a Quantitative Comparison between Experiment and Numerical Simulation}

\author[C. Gollwitzer,  G. Matthies, R. Richter, I. Rehberg, and  L. Tobiska]
{C\ls H\ls R\ls I\ls S\ls T\ls I\ls A\ls N\ns G\ls O\ls L\ls L\ls W\ls I\ls T\ls Z\ls E\ls R$^1$, G\ls U\ls N\ls A\ls R\ns  \ls M\ls A\ls T\ls T\ls H\ls I\ls E\ls S$^2$, R\ls E\ls I\ls N\ls H\ls A\ls R\ls D\ns R\ls I\ls C\ls H\ls T\ls E\ls R$^1$, I\ls N\ls G\ls O\ns R\ls E\ls H\ls B\ls E\ls R\ls G$^1$ \and L\ls U\ls T\ls Z\ns T\ls O\ls B\ls I\ls S\ls K\ls A$^3$} 

\affiliation{$^1$Experimentalphysik V, Universit\"at Bayreuth,
D-95440 Bayreuth\\[\affilskip]
$^2$ Fakult\"at f\"ur Mathematik, Ruhr-Universit\"at Bochum, Universit\"atsstra\ss e 150, 
D-44780 Bochum\\[\affilskip]
$^3$Institut f\"ur Analysis und Numerik, Otto-von-Guericke-Universit\"at Magdeburg, 
     PF 4120, D-39106 Magdeburg }

\date{\today}

\begin{document}
\maketitle

\begin{abstract}
The normal field instability in magnetic liquids is
investigated experimentally by means of a radioscopic
technique
which allows a precise measurement of
the surface topography.  The dependence of the
topography on the magnetic field is compared to results obtained by
numerical simulations via the finite element method.
Quantitative agreement has been found for the critical
field of the instability, the scaling of the pattern
amplitude and the detailed shape of the magnetic spikes. 
The fundamental Fourier mode approximates the shape to within $10\,\%$ accuracy
for a range of up to $40\,\%$ of the bifurcation
parameter of this subcritical bifurcation. 
The measured control parameter dependence of the wavenumber differs
qualitatively from analytical predictions obtained by minimization of the free energy.
\end{abstract}

\section{Introduction}
Pattern formation has  mostly  been investigated in systems driven {\em far from equilibrium}, 
like Rayleigh--B\'enard convection, Taylor--Couette flow \cite[cf.][]{cross1993}
or current instabilities \cite[]{peinke1992}. 
This lopsided orientation is partly due to the belief that mainly systems far from 
equilibrium can bring us a step forward  to comprise fundamental riddles like the origin 
of life on earth \cite[]{prigogine1988}.

However, {\em conservative} systems may also exhibit the formation of patterns, a typical example being given by 
elastic shells under a buckling load \cite[]{taylor1933,lange1971}. 
In particular, these non-dissipative systems have recently gained interest within the context of 
life: following \cite{shipman2004} they can describe pattern formation in plants.
Closely related are surface instabilities of dielectric liquids in electric fields \cite[]{taylor1965} 
and its magnetic counterpart, first observed by  \cite{cowley1967}. 
In comparison with shell structures, the surface instabilities are experimentally more accessible
because the external field can serve as a convenient control parameter. 

In order to observe the Rosensweig, or normal field instability, a horizontally extended layer of 
magnetic fluid is placed in a magnetic field oriented normally to the flat fluid surface. 
When exceeding a critical value $B_c$ of the applied magnetic induction, one can observe 
a hysteretic transition between the flat surface and a hexagonal pattern of liquid crests.
The transition gains some complexity from the fact that under variation of the surface 
three terms of the energy are varying, namely the hydrostatic energy (determined via 
the height variation of the liquid layer), the surface energy and the magnetic field energy. As the
surface profile deviates from the flat reference state, the first two terms grow 
whereas the magnetic contribution decreases. 
At the critical field, the overall energy is minimized by a static surface pattern of finite amplitude.

Using the energy minimization principle, \cite{gailitis1969, gailitis1977} and \cite{kuznetsov1976b} were able to deduce an amplitude 
equation that connects the instability to a subcritical bifurcation. However, this 
result is limited to tiny susceptibilities $\chi_0  \ll 1$. Only recently
this deficiency was partly overcome by \cite{friedrichs2001} for the normal field instability, 
and by \cite{friedrichs2002} for the more general case of the tilted field instability.
Besides these achievements, the nonlinear stability of surface patterns
under the assumption of a linear magnetization law was studied numerically 
by \cite{boudouvis1987}.

From an experimental point of view, the above predictions have been poorly investigated.
This is mainly due to the experimental difficulties in measuring the surface profiles. 
Due to their colloidal nature,  magnetic fluids are opaque and have a very poor reflectivity
\cite[]{rosensweig1985}.
They appear black to the naked eye. Thus standard optical techniques such as holography or 
triangulation \cite*[]{perlin1993} are not successful. 
Moreover, the fully developed crests are much too steep to be measured with optical 
shadowgraphy, as proposed by \cite{browaeys1999}, utilizing the slightly deformed surface as 
a (de)focusing mirror for a parallel beam of light. 
Another method, recently accomplished by \cite{wernet2001}, analyzes the reflections of a 
narrow laser beam in a Faraday experiment. However, it  yields only the local surface slope,  
but not  the local surface height. This has now been overcome by \cite{megalios2005}.
By adapting the focus of a laser beam, a height detection is possible. 
However, the method has deficiencies in accuracy, because the beam partly penetrates 
the magnetic fluid.  Moreover, it is limited to maxima and minima whereas measurements of the full surface 
topography are not possible.

\sloppypar 
A surface profile can be obtained by simple lateral observation of the instability, 
as implemented e.g.\  by \cite{mahr1998} for a single Rosensweig peak and by \cite{bacri1984}
 and \cite{mahr1998b} for a chain of peaks. It has, however, the severe disadvantage that the 
 observed crests are next to the container edge. Therefore, they are deeply affected 
 by the meniscus and field inhomogeneities. A comparison 
 with the theory for an infinitely extended layer of magnetic fluid has remained unsatisfactory for a long time.
Consequently, only the `flat aspects' of the pattern like the wavenumber
\cite*[]{abou2001,lange2000,reimann2003} or the dispersion relation \cite*[]{mahr1996,browaeys1999}
has been thoroughly investigated in experiments so far. 

Recently we have developed a radioscopic measurement technique that utilizes the attenuation of an 
X-ray beam to measure the full three-dimensional surface profile of the magnetic fluid 
far away from the container edges \cite[]{richter2001}. 

The aim of this article is to compare thoroughly the measured  surface profile of the 
instability with the analytical predictions and with novel numerical results obtained by \cite{matthies2005}
via finite element methods.

The paper is organized as follows. In the next section we give a description of
the experimental methods. Then the numerical methods of the 
corresponding simulations are explained. In the fourth section the experimental and numerical 
results are presented in parallel. Eventually, we compare and discuss these results. 

\section{Experimental methods}
In the following, we give a description of the experimental setup, the necessary image corrections,
and the calibration of the height profile. Finally we characterize the utilized magnetic liquid. 

\begin{figure}
\centering
\includegraphics[width=0.6\linewidth]{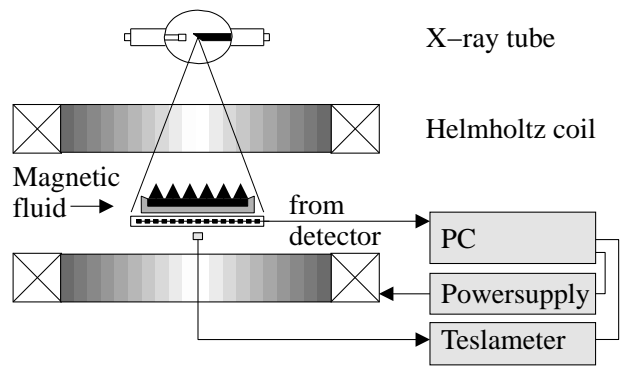}
\caption{Setup of the radioscopic measurement of the surface topography}
\label{fig:setup}
\end{figure}
We measure the surface topography of the pattern via the attenuation of X-rays passing the magnetic
fluid layer in vertical direction. Figure~\ref{fig:setup} shows the experimental setup. A container
is placed in the centre of a Helmholtz pair of coils.
The container is machined from Macrolon$^{\circledR}$ and has a diameter of $170\,$mm, and a
depth of $25\,$mm. In order to minimize the field inhomogeneity at the container edge
due to the discontinuity of  the magnetization, we have introduced a `beach'. 
The floor of the container is flat within a diameter of $130\,$mm, outside of which it is inclined
upwards at $32$~degrees,
so that the thickness of the fluid layer smoothly decreases down to 
zero towards the side of the vessel. The container is filled $10\,$mm deep with magnetic fluid.
This filling depth is in the range of the critical wavelength~($9.98\,$mm) which ensures 
according to \cite*{lange2001b} that finite size effects in vertical direction 
can be excluded.   

The two Helmholtz coils have an inner bore of $29\,\mathrm{cm}$ (Oswald Magnettechnik) and a vertical 
separation distance of $18.5\,\mathrm{cm}$. 
The field homogeneity within the empty coils is better than $0.5\,\%$ within the volume covered by the vessel.
The coils are supplied by a high precision constant current 
source (Heinzinger PTNhp 32-40) which allows to control the magnetic induction
in steps of $1\,\mathrm{\umu T}$ up to a maximum induction of $40\,\mathrm{mT}$. 

 An X-ray tube with a focus of $0.4\,\mathrm{mm} \times 1.2\,\mathrm{mm}$ 
  is mounted above the centre of the vessel at a distance of $170\,\mathrm{cm}$. 
 The tube has a wolfram anode in order to emit purely continuous radiation in the range
 of the applied acceleration voltage (20 kV to 60 kV). 
 This ensures that the beam hardness can be adapted to the absorption coefficient 
 of the fluid investigated. 
 The radiation transmitted through the fluid layer and the 
bottom of the vessel is recorded by means of an X-ray sensitive photodiode array detector. 
The detector provides a dynamic range of 16 bits, a lateral resolution of $0.4\,\mathrm{mm}$, 
and a maximum frame rate of $7.5$ pictures per second. For more details see the article by 
\cite{richter2001}.

\subsection{Image improvement}
\label{subsec:improvements}
Because of deficiencies of the X-ray detector, various digital filters had to be applied to improve
the image quality. 

First, there are so called  {\em `bad pixels'}. These are non-functional sensors which must be
ignored during image processing.
%
%
The position of those bad pixels has been extracted from an empty test image by looking for all pixels
the value of which deviates more than $10\,\%$ from the median of their neighbours. There are single isolated bad pixels as
well as few unusable rows and columns. Correction is achieved by discarding the value of the bad 
pixels and filling the gap with a bilinear interpolation from the neighbourhood. This correction of bad pixels 
can be seen as the first approximation of the Voronoi-Allebach algorithm \cite[]{sauer1987}. It is a
reasonable approach when the non-functional pixels do not form large clusters, but are rather scattered.

Second, the detector exhibits a {\em nonlinear response}, and individual pixels are not equally 
sensitive.
Let $I$ denote the real incoming intensity and $r$ the response
from the pixel, both normalized to the range $\left[0\dots1\right]$.
Then we fit the inverse response function of every individual pixel with a cubic polynomial
\begin{equation}
I(r) = a_1 r^3 + a_2 r^2 + a_3 r + a_4.
\end{equation}
The reference data for this model comes from $26$ empty test images with different illumination.


This method has an additional advantage of equalizing spatial inhomogeneities in the illumination 
caused by the nonuniformity of the intensity of the beam.
This non-uniform 
illumination is contained in the image set used for the calibration. Thus we can safely 
assume that after the nonlinear correction the value of every pixel is directly proportional to the 
absorption of X-rays at the corresponding point. The assumption, however, is that the illumination itself does not 
change over time. 

\begin{figure}
\centering
\includegraphics[width=0.7\columnwidth]{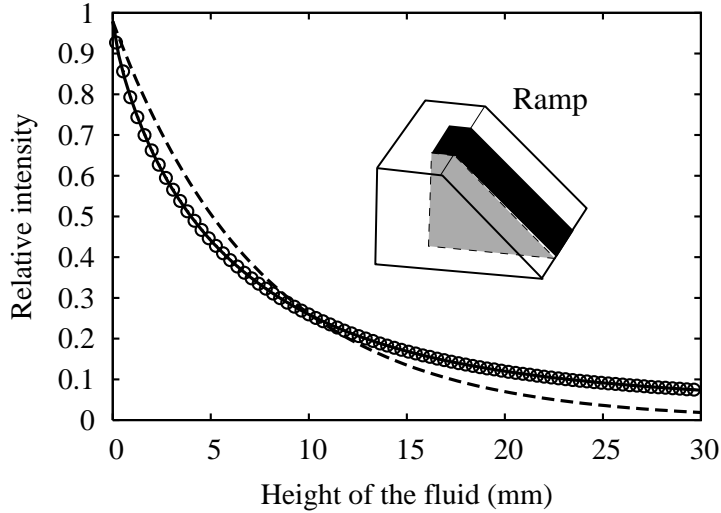}
\caption{Fluid height versus radiation intensity. The inset shows a schematic view of the wedge
filled with ferrofluid. The relative amount of the X-rays passing through the ramp can be fitted
nicely with a sum of exponential functions (solid line, equation~\ref{eq:expdecay}). A simple
exponential decay with one attenuation coefficient is not sufficient (dashed line,
equation~\ref{eq:lambertbeer}).}
\label{fig:expdecay}
\end{figure}

\begin{figure}
\centering
\includegraphics[width=\columnwidth]{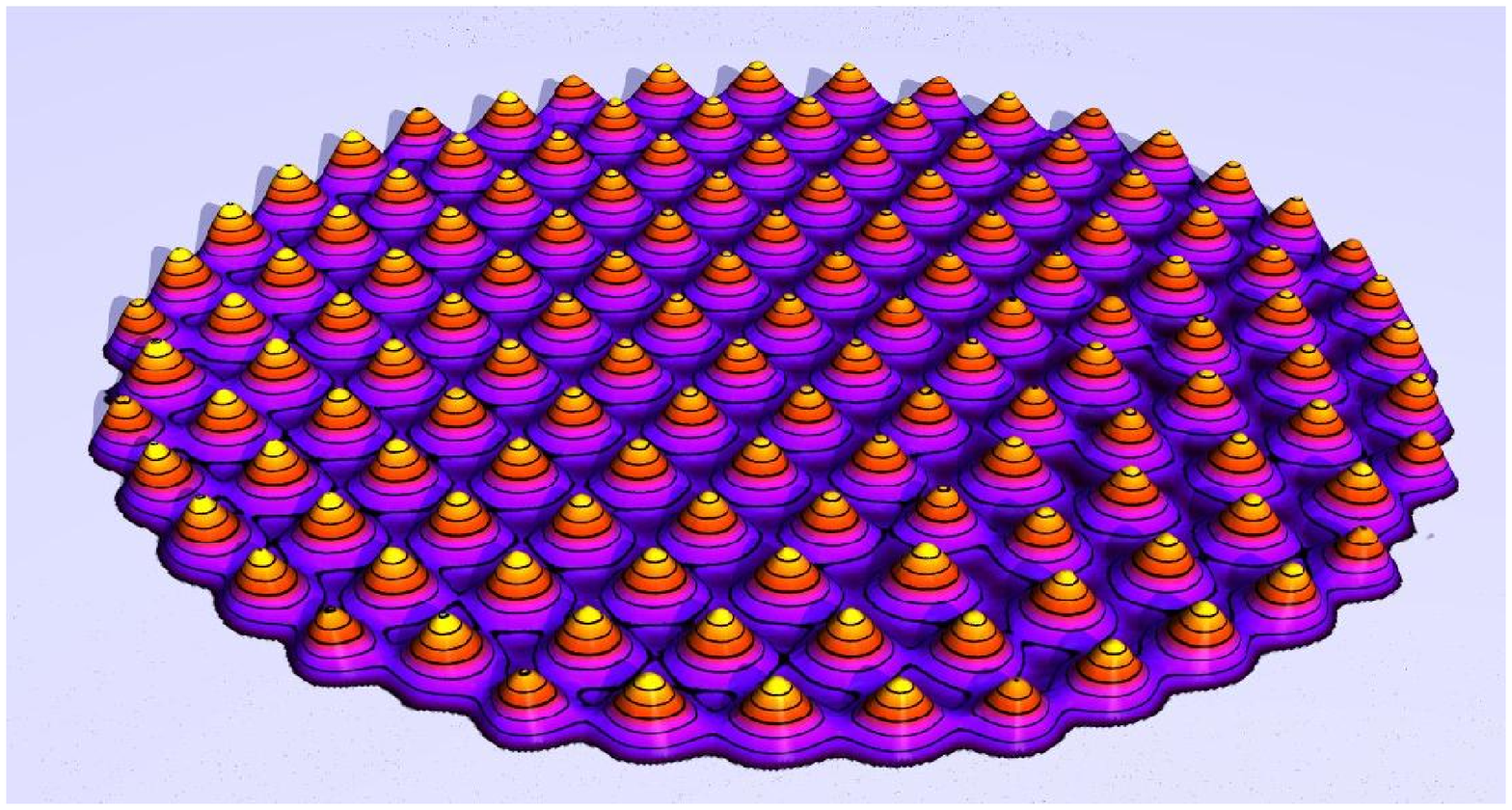}
\caption{Three-dimensional rendering of the surface pattern at $B=20.1068\,$mT. The black contour lines are $1\,$mm apart from each other. }
\label{fig:farb3d}
\vspace*{2\baselineskip}
\includegraphics[width=\columnwidth]{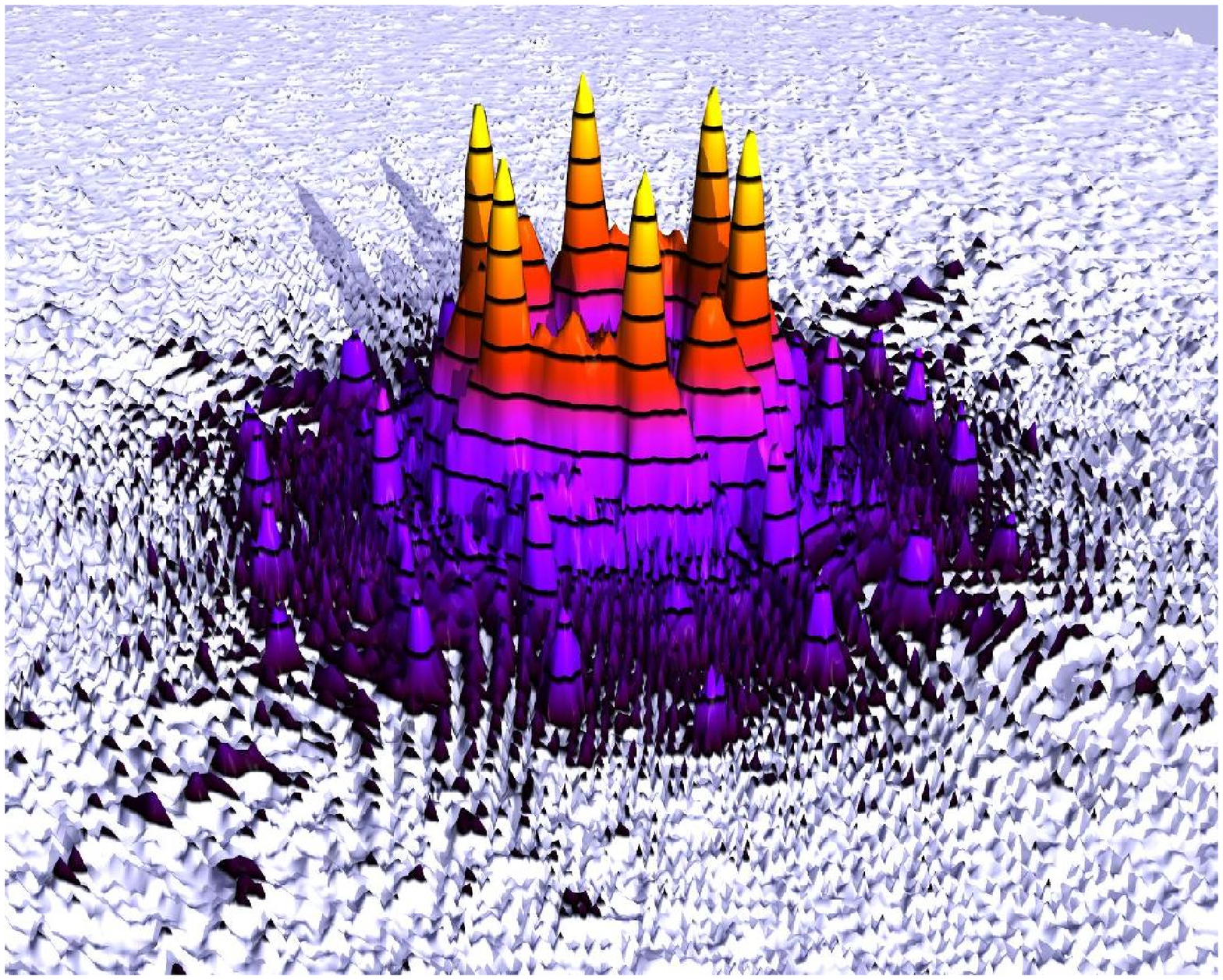}
\caption{Logarithmic Fourier space representation of the surface in figure~\ref{fig:farb3d}. The black contour lines are $6\,$dB apart from each other. }
\label{fig:farbcastle}
\end{figure}

\subsection{Conversion from intensity values to surface height profiles}
\label{sec:height}
The intensity of the X-rays decreases monotonically with the height of the fluid due to absorption.
The slope of the surface has no influence on the transmitted intensity, since X-rays have indices of
refraction very close to $1$. 

If the radiation would be monochromatic, then the decay would follow an exponential law
\begin{equation}
I(x) = I_0 \mathrm{e}^{-\beta x},\label{eq:lambertbeer}
\end{equation}
where $\beta$ is the attenuation coefficient. However, in order to get a sufficient illumination, we have to use a polychromatic
X-ray source. Hence, the absorption of the radiation depends on the wavelength and the decay cannot be
described by the simple exponential function~(\ref{eq:lambertbeer}): radiation of higher energy (`hard X-rays') is
generally  absorbed less than radiation of lower energy. In order to get a smooth approximation of the
attenuation, we fit the intensity decay with an overlay of four exponential functions, as shown in
figure~\ref{fig:expdecay}.
\begin{equation}
I(x) = I_0\,\sum_{i=1}^4 \alpha_i \mathrm{e}^{-\beta_i x}, \quad
\sum_{i=1}^4 \alpha_i = 1
\label{eq:expdecay}
\end{equation}
The datapoints were obtained by recording the absorption image of a ramp with known shape filled
with ferrofluid, see the inset in figure~\ref{fig:expdecay}. To hold the ferrofluid in position, we
covered the bottom and the side of the wedge with adhesive tape. The absorption in this tape
lowers the effective intensity of the X-rays in the fluid by about $1\,\%$, which has been taken
into account for the calculation of the absorption in the fluid. 

As can be seen
from figure~\ref{fig:expdecay}, the equation~(\ref{eq:expdecay}) is a satisfactory
interpolation method that allows us to determine the height of the fluid above every point in the image
to a resolution of up to $10\,\mathrm{\umu m}$. The absolute accuracy is not as good because of practical
problems. For example, it is difficult to determine the exact position of the wedge either
mechanically (because of its sharp end) or from the X-ray image due to the lateral resolution of $0.4\,$mm.
This means that the
absolute error of the fluid height will be around $\pm0.2\,$mm. However, when measuring the pattern
amplitude defined as a difference between the maximum and minimum surface elevation in the unit cell, this systematic
error cancels. 

After applying the corrections from the previous sections, we finally arrive at the surface profile.
A three-dimensional reconstruction of one of the recorded
profiles is shown in figure~\ref{fig:farb3d}. 
The corresponding Fourier transform is displayed in figure~\ref{fig:farbcastle}. It should be noted that
we get the Fourier space representation of the surface elevation using the radioscopic method, in contrast to other
methods where the transform of a photograph is used. This allows for interpretation of the Fourier
domain data as amplitudes which will be exploited later on in this paper.

\subsection{Tracking one single peak}
\label{sec:amplitude}
\begin{figure}
\centering
\includegraphics[width=0.6\columnwidth]{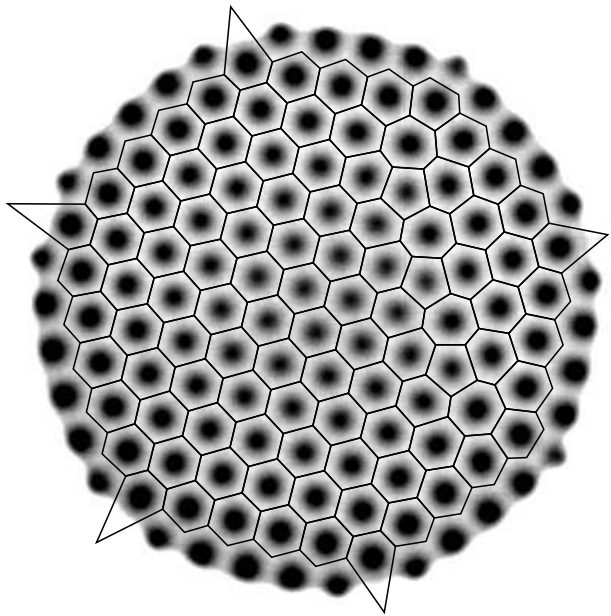}
\caption{X-Ray image and Voronoi diagram constructed from the centre of the peaks}
\label{fig:voronoi}
\end{figure}
Both analytical and numerical predictions are based on the assumption that the pattern
of peaks is periodic in space. But a perfect periodic lattice is not observed in this experiment.
For example, one can see a grain boundary between two different orientations at the right side
of the recorded profile in figure~\ref{fig:farb3d}. Additionally, as mentioned in the introduction,
peaks in the centre of the container are least affected by the edge. Thus, we select one single peak
from there. 
 Because the peak floats slightly under variation  of the magnetic induction,
it is further necessary to track it from image to image. 

To accomplish this, we first determine the position of all peaks with subpixel accuracy by fitting a
paraboloid to the centre of the peaks, and then construct the Voronoi tesselation from these positions.
The region occupied by the considered peak is then defined by its Voronoi cell, i.e.
every point in this region is closer to the considered peak centre than to any other
peak~\cite[]{fortune1995}.
We track the
movement of the considered peak on two consecutive images by finding the Voronoi cell on the later
image in which the previous position of the peak is situated. 
Figure~\ref{fig:voronoi} visualizes the tesselation. Note that the Voronoi diagram also
exposes the grain boundary made up of a chain of penta-hepta defects. 

The paraboloidal fit also yields the maximum level of this single peak. The minimum level in the
hexagonal grid is reached at the corners of the unit cell. Therefore, their arithmetic mean value is
taken as the minimum level. The amplitude of the pattern is then defined as the difference between
the maximum and minimum level.  

\subsection{Properties of the magnetic liquid}
\label{sec:matparam}
\begin{figure}
\centering
\includegraphics[width=0.8\linewidth]{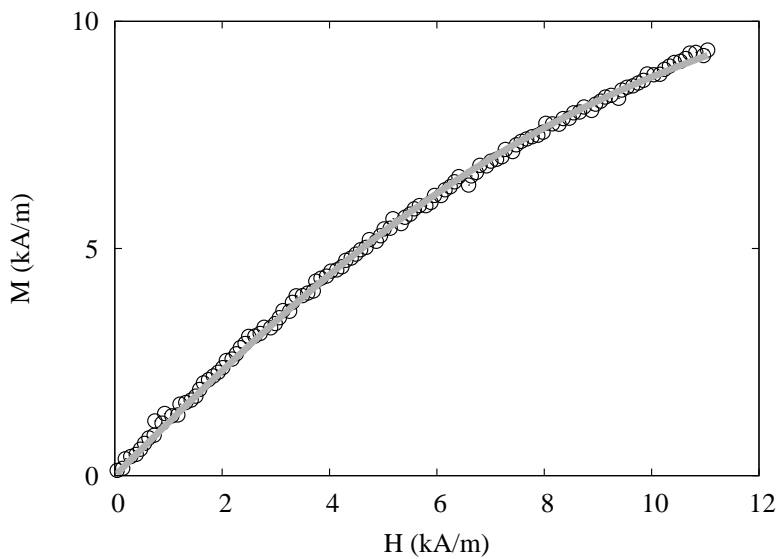}
\caption{Magnetization curve of the ferrofluid APG512a from Ferrotec Co. 
The solid line is a fit with the Langevin equation~(\ref{eq:langevin}).}
\label{fig:magnetizace}
\end{figure}

We have selected the commercial magnetic fluid APG512a (Lot~F083094CX) from Ferrotec Co.\ 
because of its convenient wavenumber and its outstanding long-term stability.
The critical field of two data series that are $5$ months apart differs by not more than $1\,\%$.
This can be attributed in part to its carrier liquid, a synthetic ester, which is commonly used 
as oil for vacuum pumps. The magnetic fluid is based on magnetite.

\begin{table}
\centering
\begin{minipage}{0.8\linewidth}
\begin{tabular}{llrl}
Quantity & & Value \phantom{$\pm$}& Error\\[3pt]
Surface tension\footnote{The absolute error
of the measurement is unknown. The error given here is taken from the analysis by
\cite{harkins1930}} &$\sigma$ & $30.57\ \pm$ & $0.1\,\mathrm{mN\,m^{-1}}$\\
Density			&$\rho$ &  $1236\ \pm$ & $5\,\mathrm{kg\,m^{-3}}$\\
Viscosity		&$\eta$  & $120\ \pm$ & $5\,\mathrm{mPa\,s}$\\
Saturation magnetization\footnote{Note that $M_S^\ast$ is not equal to the real saturation magnetization,
but is fitted in a way that equation~(\ref{eq:langevin}) approximates $M(H)$ in the initial range up to
$H=10\,\mathrm{kA\,m^{-1}}$.} &$M_S^\ast$ &  $14590\ \pm$ & $100\,\mathrm{A\,m^{-1}}$ \\
Initial susceptibility & $\chi_0$ & $1.172\ \pm$ & $0.005$
\end{tabular}
\end{minipage}
\caption{Material parameters of the magnetic fluid. }
\label{tab:matparam}
\end{table}

The characteristic properties that have an influence on the normal field instability can be found in
table~\ref{tab:matparam}. 
%
The density of the fluid has been measured using a buoyancy method and the surface tension coefficient has
been determined using a commercial ring tensiometer. 

The magnetization law $M(H)$ in the interesting range is shown in figure~\ref{fig:magnetizace}. 
Because the fluid is polydisperse, it cannot be expected that the magnetization law is a true 
Langevin function. However, fitting a Langevin function to the initial range of the magnetization 
curve leads to satisfactory results, as used before by \cite{browaeys1999}. 
The magnetization law reads therefore
\begin{equation}
\vect M(\vect H) = M_S^\ast\left(\coth(\gamma|\vect H|) - \frac{1}{\gamma|\vect H|}\right)
      \frac{\vect H}{|\vect H|} \label{eq:langevin}, \quad
\gamma = \frac{3\chi_0}{M_S^\ast}.
\end{equation}
This approximation is, of course, only valid in the initial range up to an internal field of about
$H_{\text{int}}=12\,\mathrm{kA\,m^{-1}}$. The maximum internal field in the simulation is fully 
contained within this range.

The onset of the instability can be predicted from these material parameters by the linear stability
analysis according to \cite{rosensweig1985}, \S~7.1. The critical magnetization $M_c$ of the fluid
layer is given by 
\begin{subequations}
\begin{equation}
M_c^2 = \frac{2}{\mu_0}\left(1 + \frac{1}{r_0}\right)(g\rho\sigma)^{1/2}, 
\label{eq:magkrit}
\end{equation}
where 
\begin{equation}
r_0 = \sqrt{\mu_c\mu_t}/\mu_0, \quad
\mu_c = \frac{B}{H} \quad\text{and}\quad
\mu_t = \frac{\upartial B}{\upartial H}.
\label{eq:magkrit_hilf}
\end{equation}
\end{subequations}
Together with $M(H)$ and the jump condition of the magnetic field at the ground of the dish,
\begin{equation}
B_c = H + M(H),
\end{equation}
the critical induction can be determined from these implicit equations. For the values from
table~\ref{tab:matparam}, we get $B_c=17.22\,\mathrm{mT}$.

The critical wavenumber $k_c$ from this analysis is given by
\begin{equation}
k_c =\sqrt{\frac{\rho g}{\sigma}}.
\label{eq:critwav}
\end{equation}
With our parameters, this yields $k_c= 0.629\,\mathrm{mm^{-1}}$, which corresponds to a wavelength
$\lambda_c = 9.98\,\mathrm{mm}$.


\section{Numerical methods}
Our numerical simulation of the normal field instability is based on the
coupled system of the Maxwell equations and the Navier--Stokes equations
together with the Young--Laplace equation which represents the force balance
at the unknown free interface. In this section we want to describe  a
numerical algorithm for the simulation of the coupled system of partial
differential equations. Simplified numerical models have been
studied already, see~\cite{boudouvis1987} and \cite{LMMPT03}. 
In this paper, we focus on the computation of the peak shapes 
employing a realistic magnetization curve in the form of the nonlinear Langevin function. 
This is in contrast to~\cite{boudouvis1987} where the stability of hexagonal and square pattern was considered
for a linear magnetization law.

%

We consider a horizontally unbounded and infinitely deep layer of
ferrofluid. The Maxwell equations for the non-conducting ferrofluid reduce
to
\[
\text{curl }\vect{H} = \vect{0},\qquad \text{div }\vect{B} = 0.
\]
The magnetization inside the fluid is assumed to follow equation~(\ref{eq:langevin})
while there is no magnetization outside the fluid.
The Navier--Stokes equations are given by
\begin{align}
\rho\left(
\frac{\upartial \vect{u}}{\upartial t} + (\vect{u}\cdot\boldnabla)\vect{u}
\right)
-\boldnabla\cdot \boldsymbol\tau
& = -\rho g\vect{e}_z
,\\
\text{div }\vect{u} & = 0,
\end{align}
where $\boldsymbol\tau$ is the stress tensor given by
\[
\boldsymbol\tau = \eta\left( \boldnabla\vect{u}+\boldnabla\vect{u}^T\right) 
-\left( p
+\frac{\mu_0}{2} H^2 \right)\tensor{I}
+ \vect{H}\otimes\vect{B}.
\]
Here,  $\tensor{I}$ denotes the unit tensor, $\otimes$ is the 
tensor product, 
and $p=p_\text{hyd} + p_m$ is the sum of the hydrostatic pressure 
$p_\text{hyd}$
and the fluid-magnetic pressure
\[
p_m = \mu_0\int\limits_0^H M(h)\,\mathrm{d}h.
\]
In the static case we are considering here ($\vect{u}\equiv\vect{0}$), 
the Navier--Stokes equations reduce to
\begin{equation}
\label{gradp}
\boldnabla p = -\rho g \vect{e}_z + \mu_0 M\boldnabla H.
\end{equation}
The Young--Laplace equation, which represents the force balance at the
interface, reads as follows
\begin{equation}
\label{YLeqn}
[|\boldsymbol\tau|]\vect{n} = \sigma\mathcal{K}\vect{n},
\end{equation}
where $\sigma$ is the coefficient of the surface tension, $\mathcal{K}$ the
sum of the principal curvatures and $[|\boldsymbol\tau|]$ the jump of
stress tensor.

Integrating the pressure equation~\eqref{gradp} and inserting the Young--Laplace equation~\eqref{YLeqn}, 
we obtain the relation
\begin{equation}
\label{YL2}
\sigma\mathcal{K} + \rho g z = \mu_0\int\limits_0^H M(h)\,\mathrm{d}h
+ \frac{\mu_0}{2}(\vect{M}\boldsymbol\cdot\vect{n})^2 - p_0.
\end{equation}
Here the two terms on the left-hand side characterize the surface energy
and the hydrostatic energy, while the first two terms on the right-hand side
capture the energy due to the presence of a magnetic fluid. The pressure $p_0$
in~\eqref{YL2} is a constant reference pressure.

For the numerical simulation we consider a bounded domain
$\widetilde{\Omega}\times(\widetilde{z}_b,\widetilde{z}_t)$ which is chosen in a way
that the hexagon $\widetilde{\Omega}$ contains exactly one peak. Furthermore, the
boundaries in $\widetilde{z}$-direction are assumed to be far away from the
interface. Now, the problem is transformed into its dimensionless form by
using the strength of the applied field for all magnetic quantities and a
characteristic length scale $l$ which is usually a fixed multiple of the
wavelength. The domain obtained in this way will be denoted by
$\Omega\times(z_b,z_t)$.

The Maxwell equations in dimensionless form read
\begin{equation}
\label{Maxwell}
\text{curl }\vect{H} = \vect{0},\quad \text{div }\vect{B}=0,
\qquad\text{in }\Omega\times(z_b,z_t).
\end{equation}
The first differential equation in~\eqref{Maxwell} ensures the existence of
a scalar magnetostatic potential $\varphi$ so that $\vect{H} = -\boldnabla\varphi$.
Hence, by exploiting the second differential equation of~\eqref{Maxwell},
we get
\begin{equation}
\label{magpot}
-\text{div}\big(\mu(\vect{x},|\boldnabla\varphi|)\boldnabla\varphi\big) = 0
\qquad\text{in }\Omega\times(z_b,z_t).
\end{equation}
The coefficient function $\mu(\vect{x},H)$ is given by
\[
\mu(\vect{x},H) = \begin{cases}
1 & \vect{x}\in\Omega_A,\\
\displaystyle 1 + \frac{M(H)}{H} & \vect{x}\in\Omega_F,
\end{cases}
\]
where $\Omega_F$ and $\Omega_A$ are the subdomains of $\Omega\times(z_b,z_t)$
which correspond to the regions inside and outside the ferrofluid, respectively.
The magnetostatic problem~\eqref{magpot} is a nonlinear uniformly elliptic partial
equation. The nonlinearity in~\eqref{magpot} is resolved by a fixed-point
iteration. The partial differential equation~\eqref{magpot} is equipped with the
following boundary conditions:
$\varphi = -z H^{(F)}$ at the bottom boundary
$z_b$, $\varphi = -z H^{(A)}$ at the top boundary $z_t$, and
$\upartial\varphi/\upartial n = 0$ at the side boundary. Here, $H^{(A)}$ and
$H^{(F)}$ denote the constant strength of the magnetic field outside and inside
the ferrofluid in the case of an undisturbed interface, respectively.
Moreover, $H^{(F)}$ can be obtained from  $H^{(A)}$ by a single algebraic equation.
    
In the consideration of the Young--Laplace equation we assume that the
interface $\Gamma$ is the graph of a function $\Psi:\Omega\to\mathbb{R}$,
i.e.,
\[
\Gamma = \big\{ (x,y,z)\in\mathbb{R}^3\::\: z= \Psi(x,y),\:(x,y)\in\Omega\big\}.
\]
Now, the sum of the principal curvatures can be written in terms of $\Psi$. 
We
have
\[
\mathcal{K} = -\text{div}\frac{\boldnabla \Psi}{\sqrt{1+|\boldnabla \Psi|^2}}.
\]
Hence, after dividing by $\sigma$, the Young--Laplace equation is given by
\begin{equation}
\label{YL3}
-\text{div}\frac{\boldnabla \Psi}{\sqrt{1+|\boldnabla \Psi|^2}} + \Lambda^2 \Psi = F
\qquad \text{in }\Omega,
\end{equation}
where $F$ contains all magnetic terms and $\Lambda$ is the critical wavenumber 
$k_c$ expressed in units of the inverse characteristic length scale $l^{-1}$. 
The differential equation is
completed by the boundary condition $\upartial \Psi/\upartial n=0$ due to
symmetry. The Young--Laplace equation is a nonlinear elliptic equation which
is, however, non-uniformly elliptic. The nonlinearity is again resolved by a
fixed-point iteration.

Both the magnetostatic problem~\eqref{magpot} and the Young--Laplace
equation~\eqref{YL3} are solved
approximately by finite element methods. The magnetostatic problem is a
three-dimensional equation which is discretized by continuous piecewise
trilinear functions on hexahedra. The Young--Laplace equation is a
two-dimensional problem. For its discretization, continuous piecewise
bilinear functions on quadrilaterals are used.

\begin{figure}
\parbox{0.4\columnwidth}{%
\centering
\includegraphics[width=\linewidth]{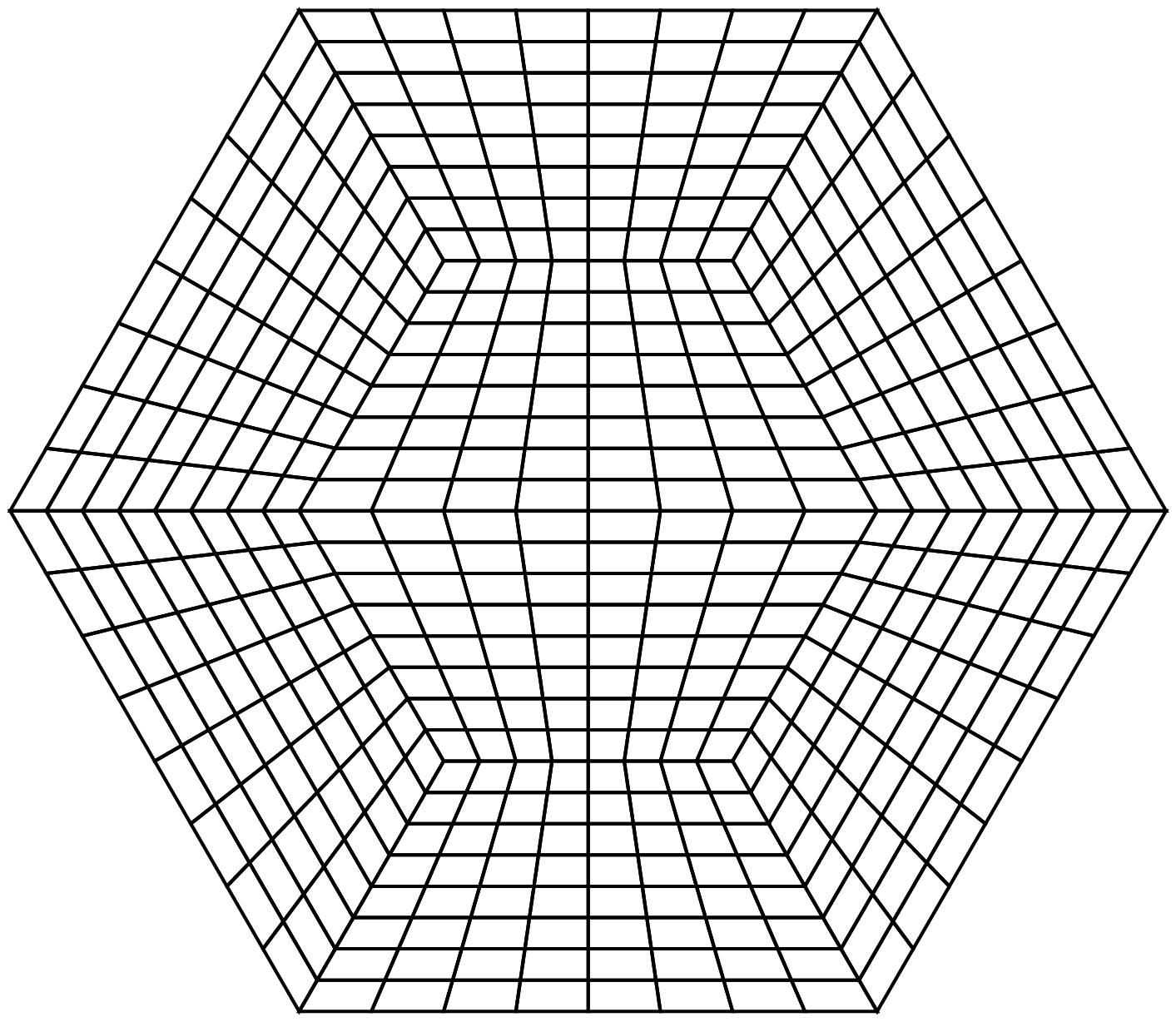}\\
(a)}\hfill
\parbox{0.45\columnwidth}{%
\centering
\includegraphics[width=\linewidth]{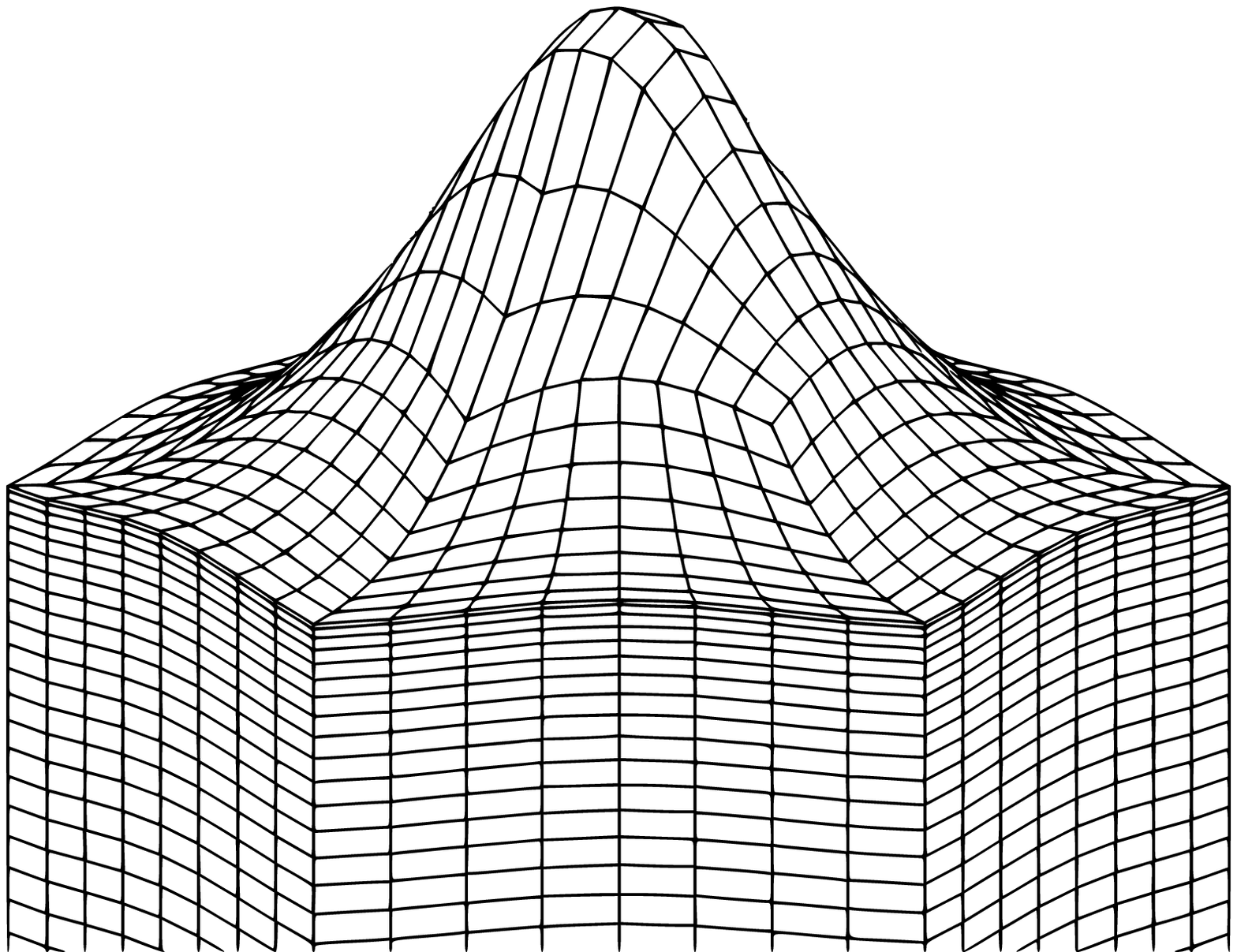}\\
(b)}
\caption{Two-dimensional mesh (a) and three-dimensional surface mesh (b).\label{fig:meshes}}
\end{figure}

Figure~\ref{fig:meshes} shows a mesh for the Young--Laplace equation and
a three-dimensional surface mesh on the peak.

We have to solve two large systems of nonlinear algebraic equations which
correspond to the three-dimensional problem for the magnetostatic potential $\varphi$
and the two-dimensional problem for the function $u$ which describes the unknown free
surface. In both cases, fixed-point iterations have been applied while the
arising linear systems of equations were solved by a multi-level algorithm in
each step of the iteration. In the three-dimensional problem for the magnetostatic potential,
a geometric multi-level method has been applied based on a family of
successively refined three-dimensional hexahedral meshes. The Young--Laplace equation is
solved on quadrilateral mesh which is the projection of the three-dimensional surface mesh
onto a plane. The arising two-dimensional problems were solved by an algebraic (instead
of a geometric) multi-level method. The coupling of three-dimensional and two-dimensional problems results
in quite difficult data structures which are needed for the information
transfer between the subproblems. These data structures had to be
developed and were implemented in the program package MooNMD~\cite[]{JM04}.

\begin{figure}
\begin{center}
\includegraphics[width=0.5\columnwidth]{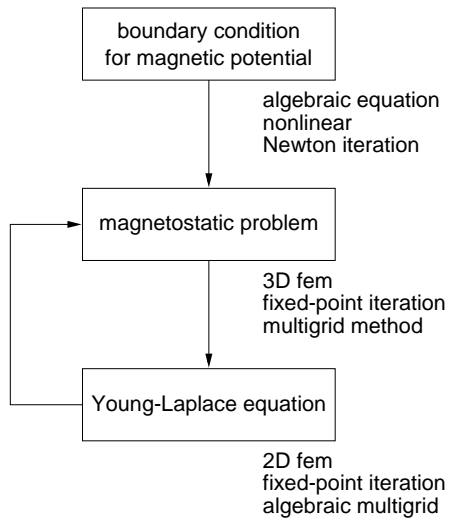}
\caption{Flow chart.\label{fig:flow}}
\end{center}
\end{figure}

All iteration processes are illustrated in the flow chart shown in
figure~\ref{fig:flow}.

\begin{table}
\centering
\begin{tabular}{lrrr}
Level & \multicolumn{1}{c}{2} & \multicolumn{1}{c}{3} &
\multicolumn{1}{c}{4}\\[3pt]
d.o.f. (Young--Laplace equation.) & 141 & 537 & 2\,097\\
d.o.f. (magnetic potential) & 6\,909 & 52\,089 & 404\,721
\end{tabular}
\caption{Number of unknowns (degrees of freedom = d.o.f.) on different
refinement levels.\label{tab:dof}}
\end{table}

Table~\ref{tab:dof} shows that the greatest computational costs are caused by
the solution of the magnetostatic problem~\eqref{magpot} since the three-dimensional
problems have many more unknowns than the associated two-dimensional problems.
Note that the
number of unknowns for the three-dimensional problem increases by a factor of~8 in one
refinement step while the number of unknowns for the two-dimensional problem increases
only by a factor of~4.

We carried out numerical simulations with the material parameters given in table~\ref{tab:matparam}.
\begin{figure}
\begin{center}
\includegraphics[angle=270,width=0.9\columnwidth]{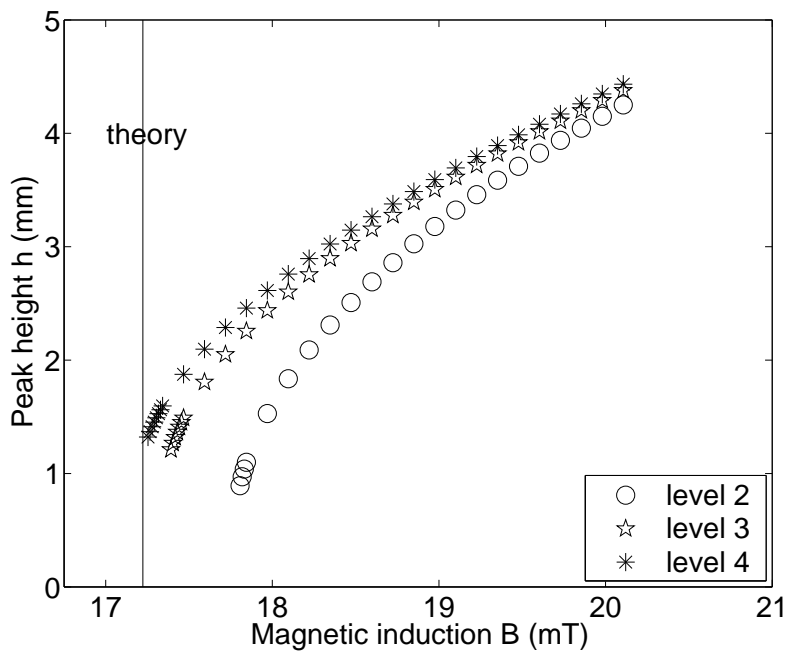}
\caption{Peak height depending on the applied field strength for different
refinement levels.\label{difflevel} The thin vertical line represents the
critical induction of the instability (from the linear theory, see \S~\ref{sec:matparam}).}
\end{center}
\end{figure}
Figure~\ref{difflevel} shows the peak height as a function of the applied
field for different refinement levels of the mesh for the underlying finite
element method. The given peak height is the difference between the highest point on the surface,
i.e. at the mid-point of a hexagonal cell, and the lowest point, i.e. at one of the corners of the
hexagonal cell.
Furthermore, the theoretical value for the onset of the instability is
shown in Figure~\ref{difflevel}. We see that the qualitative behaviour is
reproduced even on very coarse meshes. Obviously, one gets higher
peaks on finer meshes. Moreover, we obtain numerically on the finest
considered mesh a value for the critical magnetic induction which is very
close to the theoretical value.
\section{Results}
In the experiment we recorded $540$ surface profiles in total, increasing and decreasing the magnetic induction in a
quasistatic manner from $16.7\,\mathrm{mT}$ to $20.1\,\mathrm{mT}$ at maximum in steps of
$0.015\,\mathrm{mT}$. Below this range, the surface remains flat apart from attraction of the fluid
to the boundary of the container. Above this range the hexagonal pattern transforms into a square array of
peaks which is not considered in the present paper. 

In the next paragraphs, we will compare characteristic properties from these profiles to their
counterpart from the simulated peaks. First we compare the amplitude of the pattern. Then we look at
the wavenumber, and finally the full shape is examined.

\subsection{Scaling behaviour of the amplitude}
\label{sec:scaling}
\begin{figure}
\centering
\includegraphics[width=0.8\columnwidth]{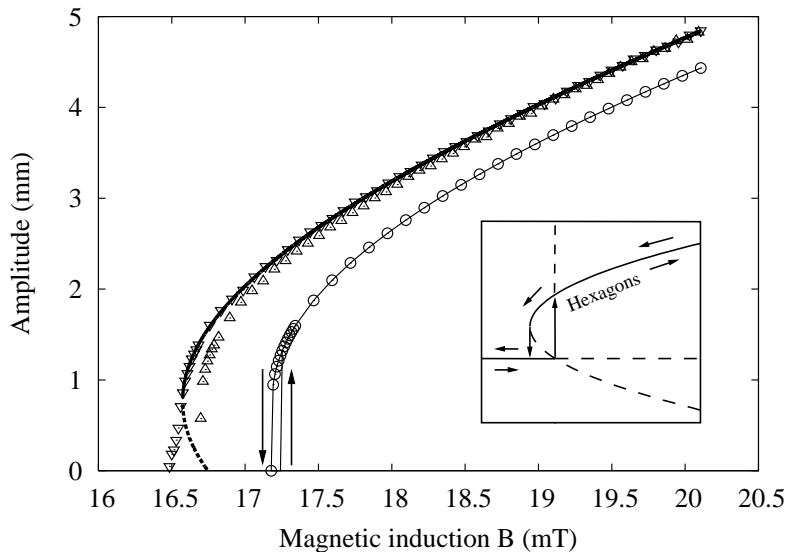}
\caption{Pattern amplitude as a function of the magnetic induction. Triangles pointing upwards~(downwards) represent the 
experimental data for an increasing~(decreasing) magnetic field. The symbol size approximates the
statistical error. For clarity only every 5th point is plotted.
The connected circles show the simulation data. The thin vertical lines represent the numerically
found transition points from~(to) the flat surface. 
The thick solid line is a fit of the experimental data
with the solution of an amplitude equation, the dashed line is representing the unstable branch of the solution. 
The inset reproduces the qualitative behaviour expected from the analysis by \cite{gailitis1977}.}
\label{fig:linbifurk}
\end{figure}

Figure~\ref{fig:linbifurk} shows the amplitude of the pattern as a function of the applied magnetic
induction, as defined in section~\ref{sec:amplitude}. 
The triangles denote the experimental values where the size of the symbols has been chosen
to approximate the statistical error of the data. The open circles represent the result from the simulations.
The solid line is a fit of the experimental data with the solution of an amplitude equation from \cite{friedrichs2001}.
The root of the amplitude equation  reads
\begin{equation}
A(\varepsilon) k_c= \frac{1}{2b_1}\left(b_2(1+\varepsilon)\pm 
  \sqrt{b_2^2(1+\varepsilon)^2+4\varepsilon b_1)}\right), 
  \label{eq:amplitude}
\end{equation}
where $k_c = 0.629\,\mathrm{mm^{-1}}$
is the critical wavenumber of equation~(\ref{eq:critwav}) and $\varepsilon = (B^2-B_c^2)/B_c^2$ is the bifurcation
parameter. The fit parameters are
\begin{align}
B_c   & =  16.747 \pm 0.001\,\mathrm{mT},\label{eq:critinduction}\\
b_1     & = 0.0889\pm 0.0001, \\
b_2 & = 0.0873 \pm0.0003. 
\end{align}
Using these parameters, the analytical function describes the measured data very well. However, it contains three
adjustable parameters. The coefficients computed by \cite{friedrichs2001} are not applicable with
our material parameters, because the expansion is only valid up to a susceptibility $\chi_0=1.05$.
So the analytical theory has no predictive power for our fluid.

Let us now compare the measured amplitude with the results from the simulation which does not use a
single adjustable parameter.
Qualitatively, the numerical and experimental data compare very well. Both curves show a small
bistable range, which makes it necessary to control the magnetic
field in tiny steps to resolve the hysteresis. In the experiment, the width of the hysteresis loop
is $0.17\,\mathrm{mT}$ whereas the numerical data expose a hysteretic range of $0.06\,\mathrm{mT}$. 
For a higher concentrated fluid ($\chi_0=2.2$), we have recently observed a larger hysteretic range of $1\,\mathrm{mT}$
using the same setup \cite[]{richter2005}. This indicates that for even smaller susceptibilities
additional care has to be taken to resolve any hysteresis. 

For both experimental and numerical data, the amplitude jumps at the critical point to a
height of about $1.5\,\mathrm{mm}$ and reaches about $5\,\mathrm{mm}$ at the highest field. When decreasing the field,
the surface pattern vanishes at the induction $B_\ast < B_c$ with a sudden drop from about $1\,$mm.
Despite the principal similiarity of experimental and numerical results, the agreement is not
convincing: at corresponding amplitudes the experimental data are shifted to lower fields.

\begin{figure}
\parbox{0.25\columnwidth}{%
\centering
\includegraphics[width=\linewidth]{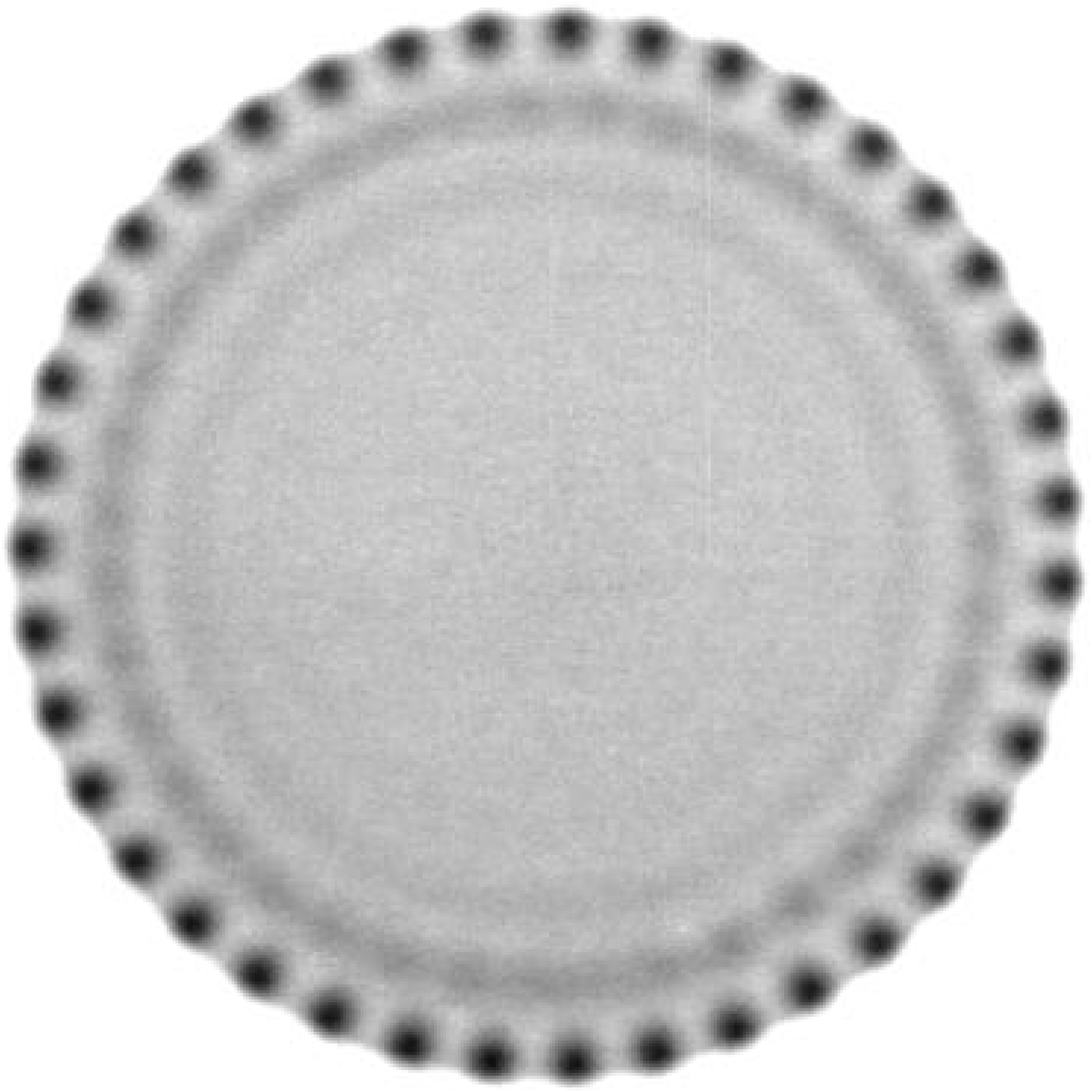}\\
(a)}\hfill
\parbox{0.25\columnwidth}{%
\centering
\includegraphics[width=\linewidth]{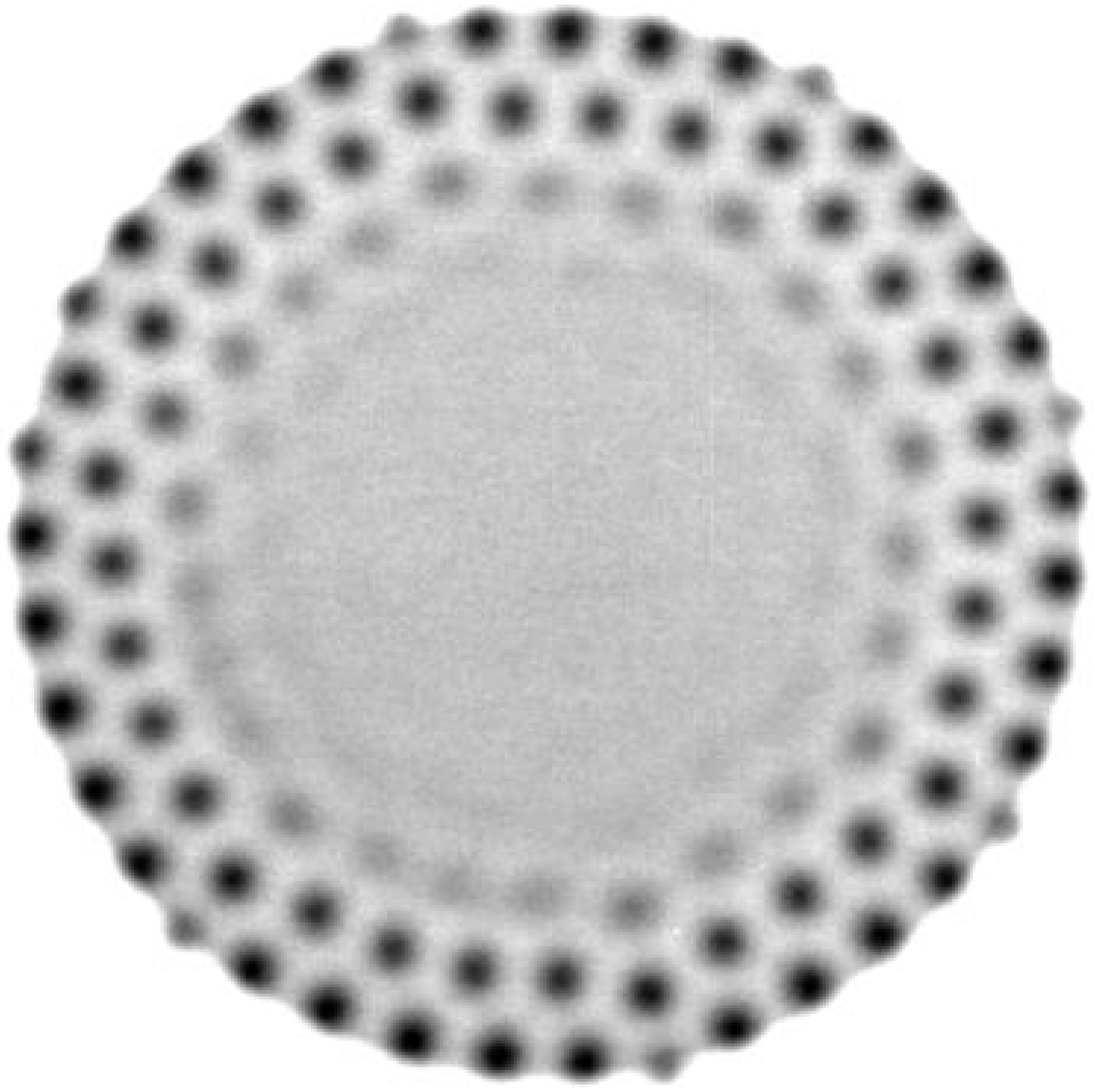}\\
(b)}\hfill
\parbox{0.25\columnwidth}{%
\centering
\includegraphics[width=\linewidth]{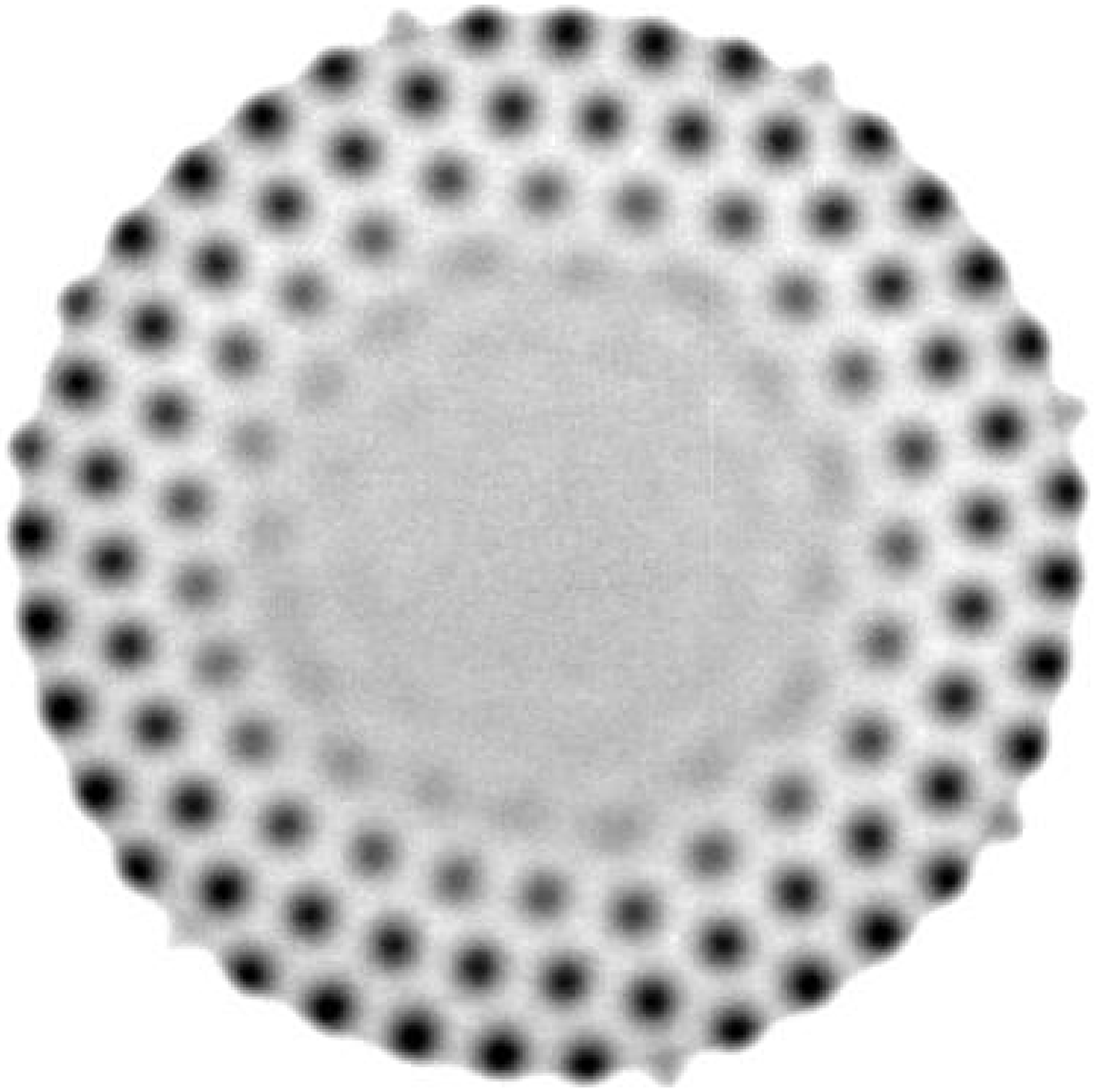}\\
(c)}\\[\baselineskip]
\parbox{0.25\columnwidth}{%
\centering
\includegraphics[width=\linewidth]{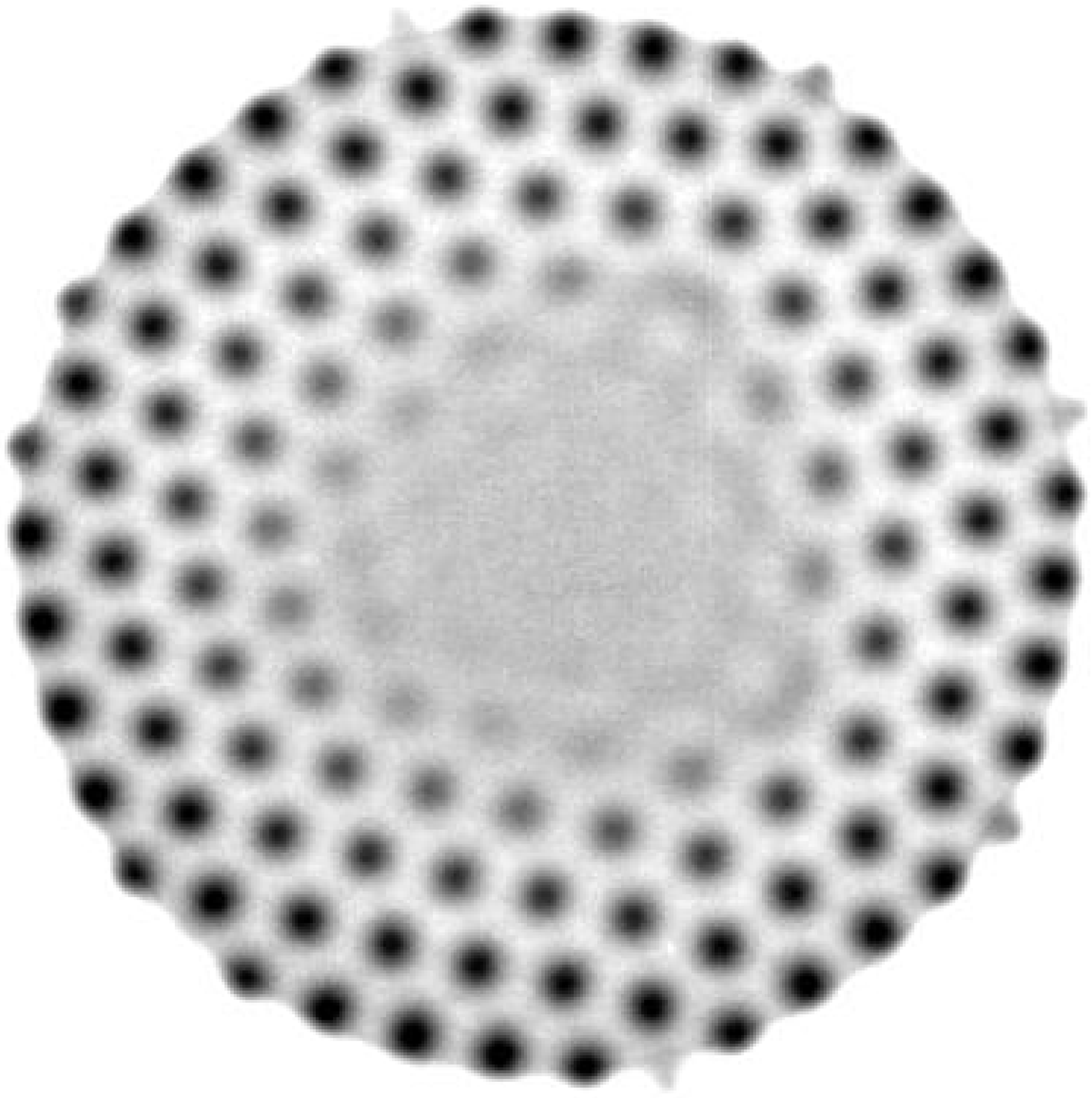}\\
(d)}\hfill
\parbox{0.25\columnwidth}{%
\centering
\includegraphics[width=\linewidth]{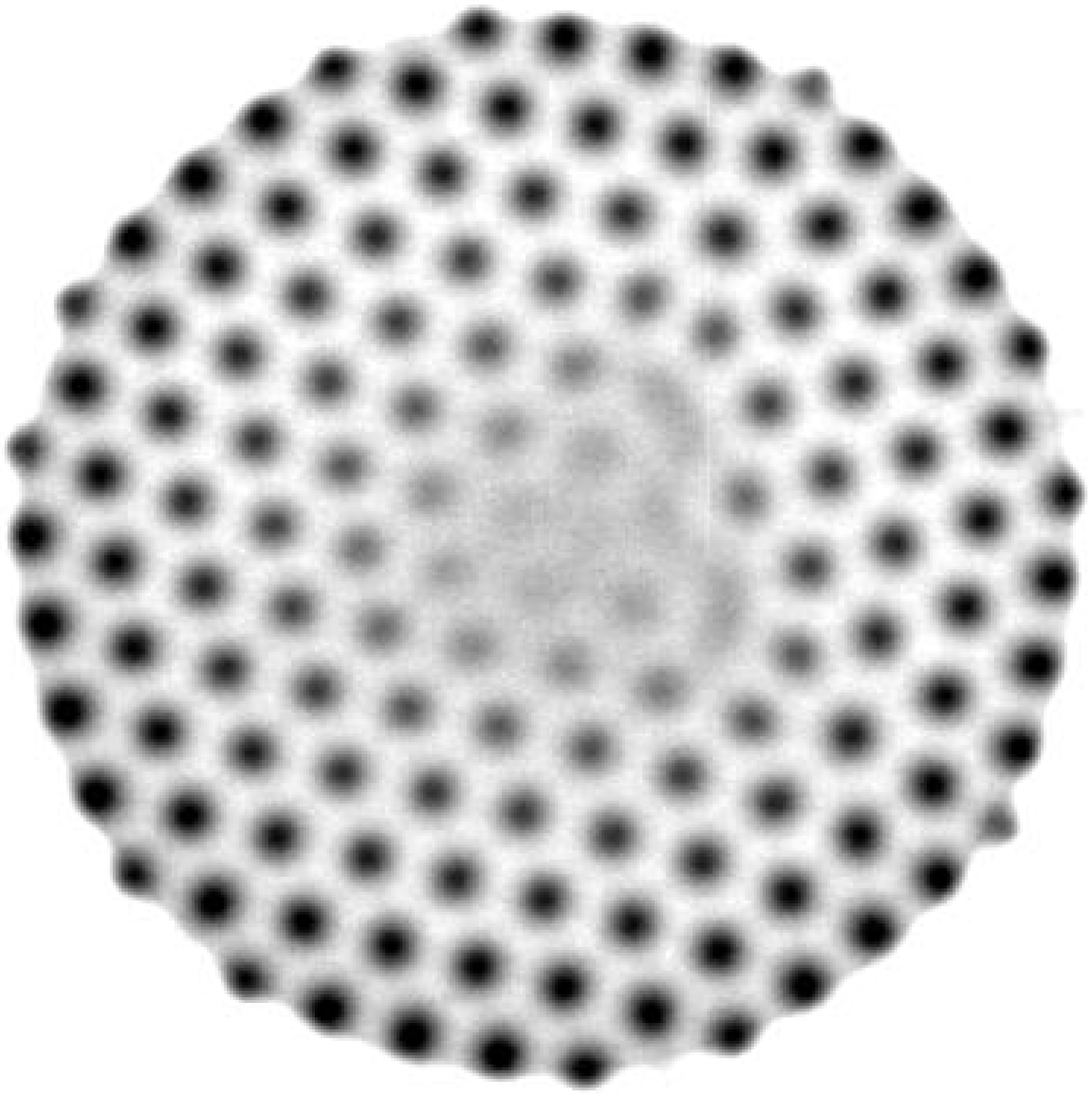}\\
(e)}\hfill
\parbox{0.25\columnwidth}{%
\centering
\includegraphics[width=\linewidth]{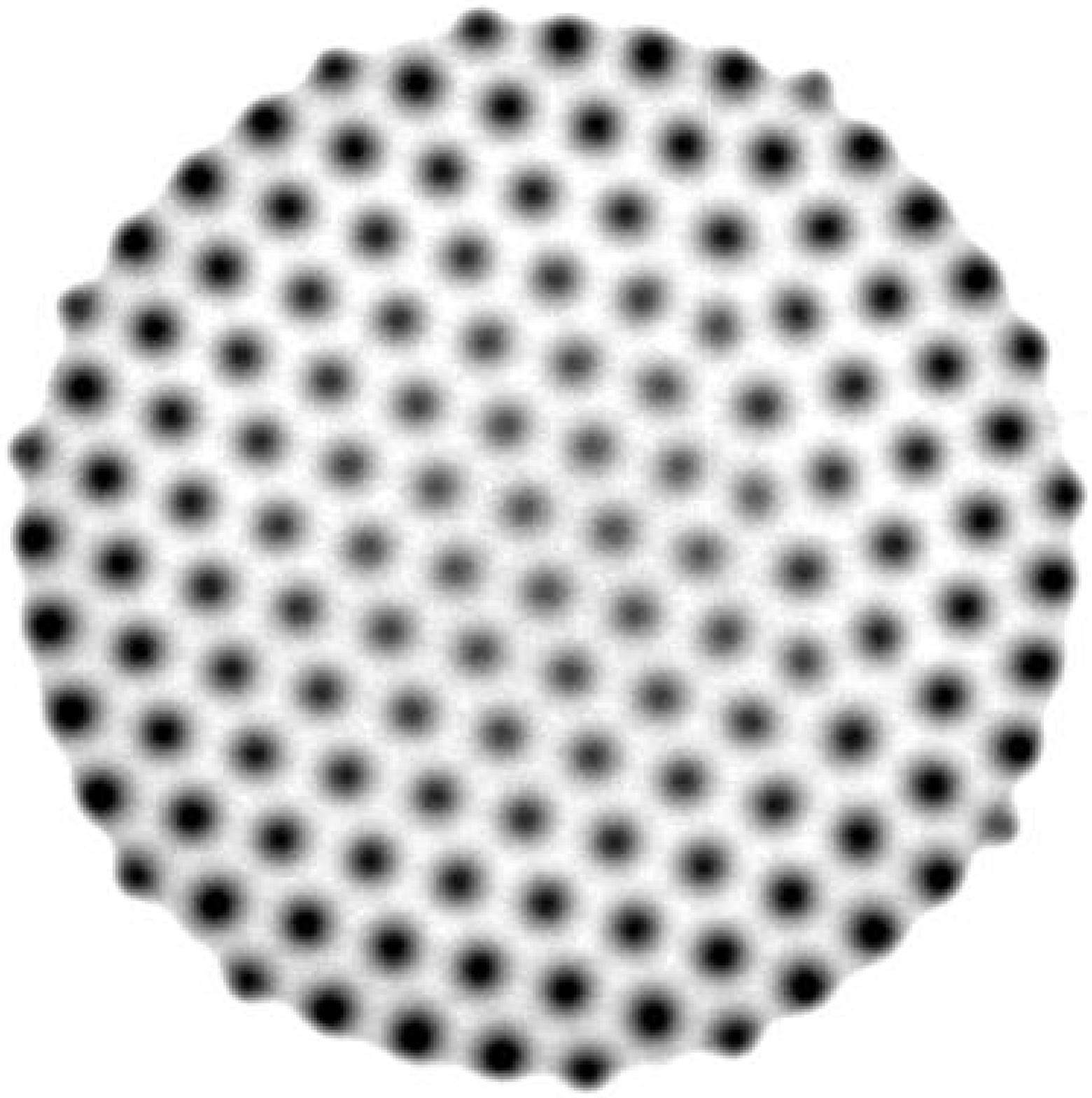}\\
(f)}

\caption{The pattern emerging as the magnetic induction is increased. The induction at the centre of
the vessel is
(a)~$B=16.45\,\mathrm{mT}$,
(b)~$B=16.51\,\mathrm{mT}$,
(c)~$B=16.56\,\mathrm{mT}$,
(d)~$B=16.62\,\mathrm{mT}$,
(e)~$B=16.68\,\mathrm{mT}$,
(f)~$B=16.73\,\mathrm{mT}$.}
\label{fig:imperfection}
\end{figure}

The critical induction seen in the simulations $B_c^\text{sim} = 17.25\,\mathrm{mT}$, i.e.\ the induction at which the jump occurs when
increasing the field, is in accordance with the theoretical value from the linear stability analysis
$B_c = 17.22\,\mathrm{mT}$, see \S~\ref{sec:matparam}. From the experimental data, the critical field can be
extracted by the fit with equation~(\ref{eq:amplitude}), which yields $B_c = 16.747\,\mathrm{mT}$. It
deviates only by $3\,\%$ from the theoretical value. The
difference between these two thresholds does not lie within the statistical error, which indicates that it
is a systematic deviation: with the errors given in section~\ref{sec:matparam}, the uncertainty of
the theoretical value $B_c$ is about $1.1\,\%$.

The imperfection induced by the edge of the bounded container has a great influence over the emerging
pattern. Figure~\ref{fig:imperfection} shows six consecutive X-ray images for increasing magnetic induction.
Because of the inhomogeneous magnetic induction over the vessel, pattern formation starts at the
edge and expands towards the centre until the whole surface is covered with peaks.  

\begin{figure}
\centering
\includegraphics[width=0.8\columnwidth]{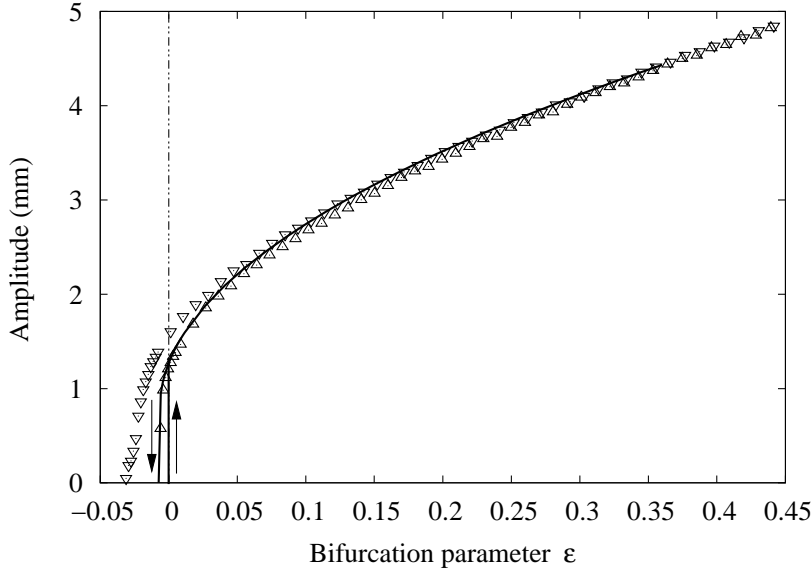}
\caption{Pattern amplitude as a function of the bifurcation parameter. 
$B_c$ has been chosen as 16.747~mT for the experimental data (from
equation~\eqref{eq:critinduction}) and 17.22~mT for the numerical data (from the linear theory). The
simulation data are shown as a solid line only, whereas the experimental data are represented by the
upward~(downward) triangles for increasing~(decreasing) magnetic field.}
\label{fig:epsbifurk}
\end{figure}

Nevertheless, the simulations can be reconciled with the experimental findings if we take
the shift in the critical field into account. To see this, we plot both curves in a unifying diagram,
see figure~\ref{fig:epsbifurk}.  Instead of the magnetic induction $B$, we plot the amplitude as a function
of the bifurcation parameter $\varepsilon = (B^2-B_c^2)/B_c^2$. Here, the actual $B_c$ is used for each data
set.
Now, the simulation data (plotted as a solid line only) match nearly perfectly the experimental
data represented by the symbols. Slight deviations can be found near $\varepsilon=0$: for
$\varepsilon>0$, the experimental amplitudes for decreasing and increasing field differ while the
simulation produces identical results which fall somewhere in between these two curves.  For
$\varepsilon_\ast<\varepsilon<0$, i.e.\ in the range where the surface is bistable, the experiment
already shows a small-amplitude surface pattern for increasing induction which should not be there in the ideal case.

The range of bistability observed in the experiment is about twice as wide as the one found in the simulation.
Since the transition is not equally sharp due to the imperfection, it is not easy to determine 
the exact range where the surface is bistable.

\subsection{The wavenumber modulus}
\begin{figure}
\centering
\includegraphics[width=0.8\columnwidth]{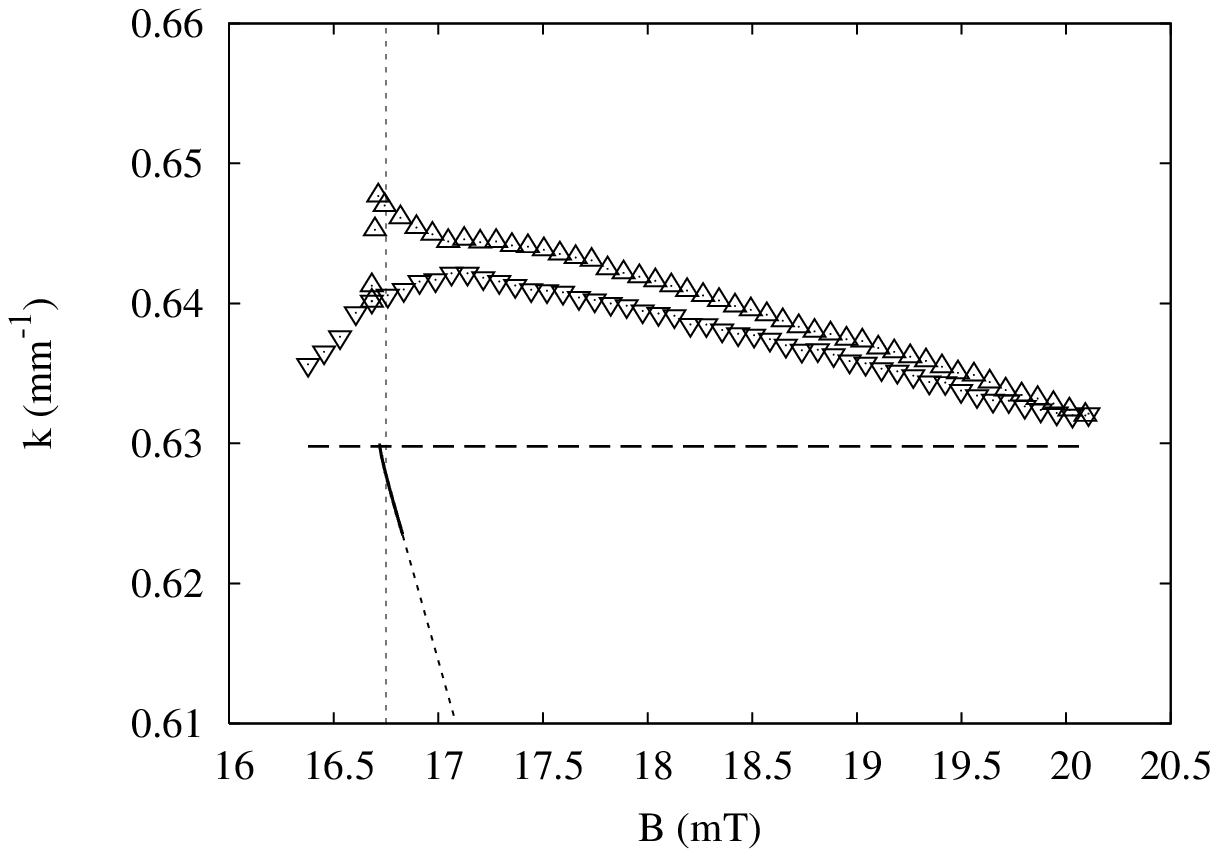}
\caption{Wavenumber of the experimental peak pattern, as determined from Fourier space. Triangles
pointing downwards designate decreasing magnetic induction, upward triangles stand for rising
induction. The dotted vertical line represents the critical induction from equation~\eqref{eq:critinduction},
the dashed line marks the critical wavenumber used in the simulations.
The solid line is the preferred wavenumber taken from the theory by \cite{friedrichs2001}, but
computed for a different set of parameters (particularly, $\chi_0=0.35$), the linear extrapolation
is dotted. }
\label{fig:wavenumber}
\end{figure}
Figure~\ref{fig:wavenumber} shows the experimental wavenumber
for increasing and decreasing magnetic field which has been determined in the Fourier space
with high precision (figure~\ref{fig:farbcastle}). Subpixel accurracy can be achieved due to the fact,
that the Fourier space representation is a direct transform of the surface topography, not the transform 
of a flat photograph. 
Thus it
includes the correct amplitudes. As the magnetic induction $B$ is increased, the wavenumber first
increases and then decreases slowly. When $B$ is reduced afterwards,  $k$ increases again, but
to slightly smaller values; hence, $k$ exhibits a small hysteresis loop.   

\cite{friedrichs2001} predict that the critical wavenumber of maximal growth $k_c$ should always
be larger than the wavenumber of the resulting nonlinear pattern. 
However, in our experiment $k_c$ was found to be smaller than the wavenumber of the pattern, 
with the difference of the two being less than $3\,\%$. 
Further, the computed wavenumber decreases monotonically, as the
magnetic induction is increased. In our experiment, instead, it has a maximum near the critical point.
This also contradicts the findings of \cite{bacri1984} and \cite{abou2001}, which report a constant wavenumber.
Note that the wavenumber measured here should not be
confused with the wavenumber of maximal growth which is predicted by a linear stability analysis.
This latter wavenumber shows a linear increase with $B$ as measured by \cite{lange2000}. During the
nonlinear stabilization of the pattern, the wavenumber relaxes to some other value that is induced
by the boundary conditions \cite[]{lange2001}. In all previous experiments the container had straight vertical
edges corresponding to hard boundary conditions, which forces the wavenumber to be an integer
multiple of the reciprocal container diameter. In contrast to that we equipped our container with a
ramp that should give more freedom to the wavenumber. Such a ramp has been studied extensively e.g.\
in the context of convection by \cite{rehberg1987}. It is obvious from figure~\ref{fig:wavenumber} that
our ramp permits different wavenumbers for the same magnetic induction. Nonetheless the boundary seems to
provide a soft pinning effect, which 
selects a wavenumber that is not necessarily the preferred one computed by \cite{friedrichs2001}.

Because of the rather small difference between the experimentally found wavenumber and $k_c$, 
the wavenumber modulus $k$ has been fixed to the critical value
$k_c = 0.629\,\mathrm{mm^{-1}}$ (cf.\ \S~\ref{sec:matparam}) in the simulations. In principle, a numerical ab initio estimation of the
wavenumber of the pattern for each value of the induction is possible. This involves calculating the
surface pattern for different preset wavelengths. Afterwards the preferred wavenumber can be
selected by the minimum of the total free energy of the simulated profiles. However, the computational cost of this technique was too high at the present stage.
Since the deviation of
the experimental wavenumber from the critical one is only about $3\,\%$ at maximum, it 
is expected that the simulated profiles are a near match of the experiment.

\subsection{The shape of the peaks}
\begin{figure}
\parbox{0.49\linewidth}{%
\centering
\includegraphics[width=\linewidth]{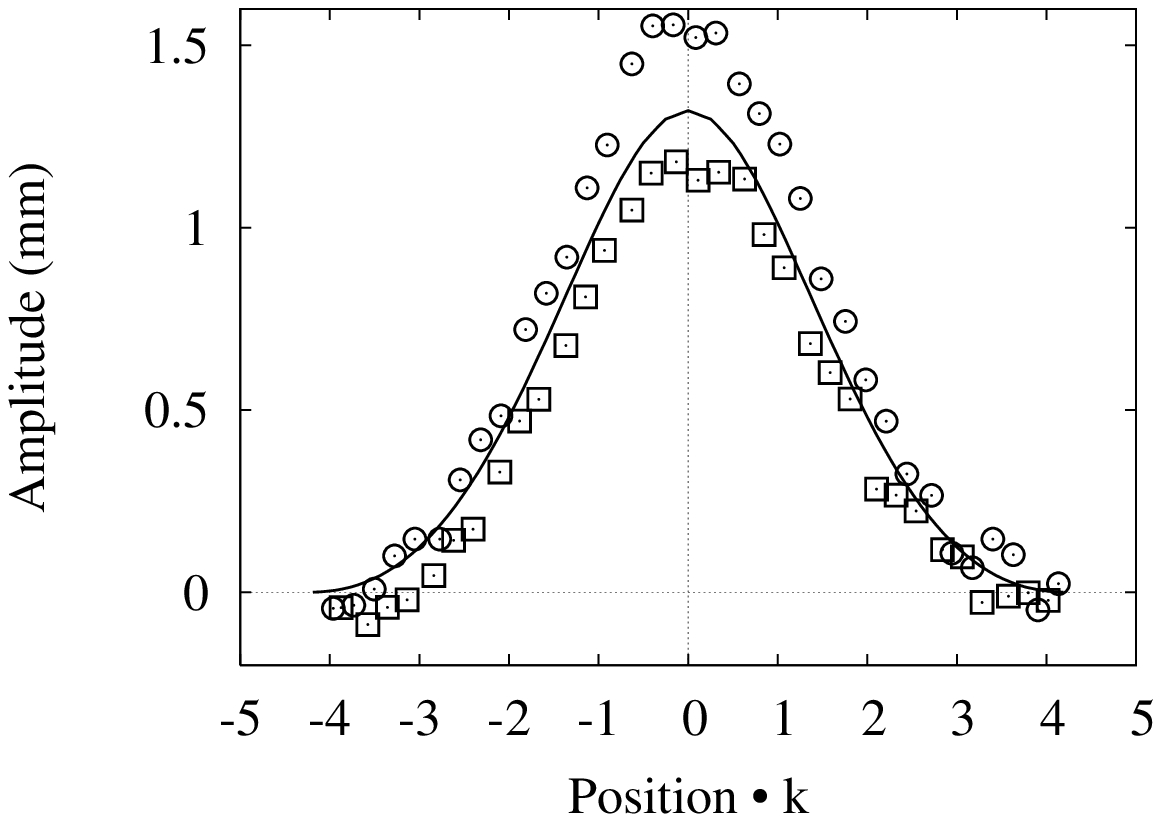}\\
(a)}\hfill
\parbox{0.49\linewidth}{%
\centering
\includegraphics[width=\linewidth]{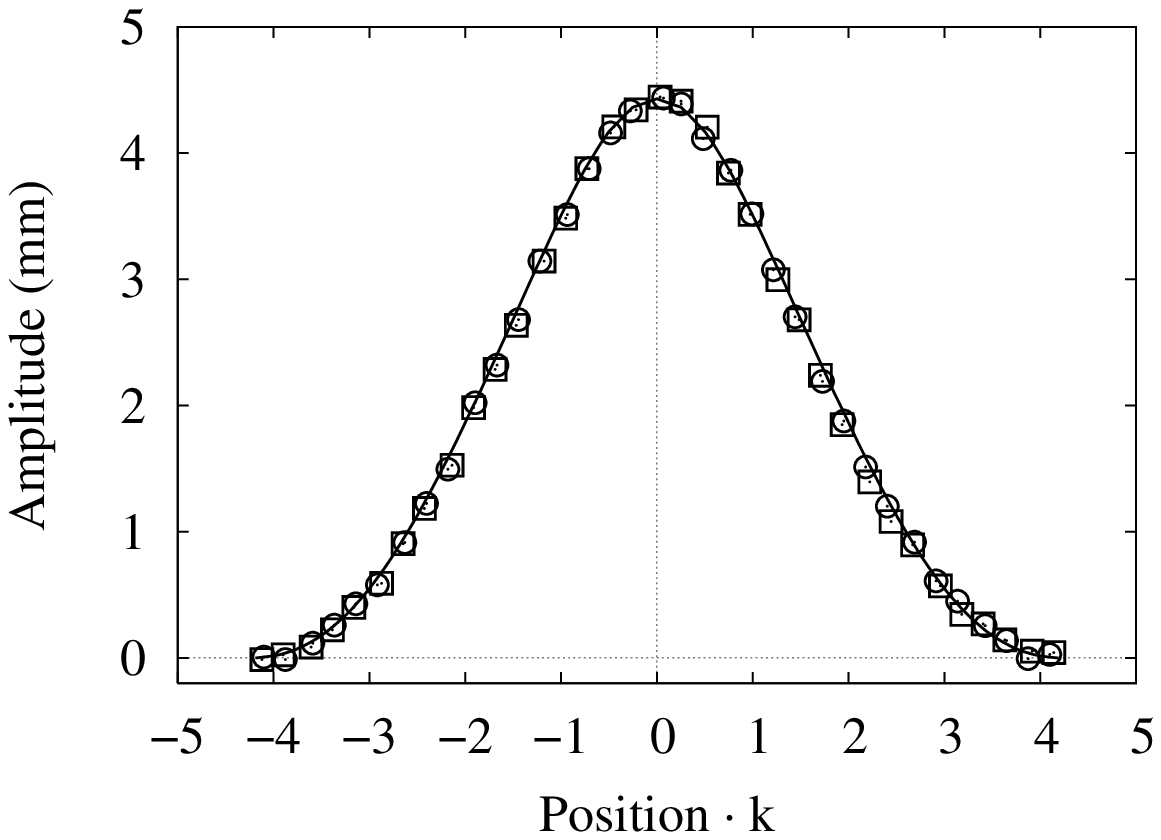}\\
(b)}
\caption{Comparison of experimentally (symbols) and numerically (solid lines) obtained peak profiles at 
$\varepsilon=0$~(a) and $\varepsilon=0.35$~(b)}
\label{fig:profile}
\end{figure}

Let us now consider the shape of the liquid crests. Figure~\ref{fig:profile} presents a direct comparison
of the measured profile of a single peak selected from the centre of the dish with the shape of the peak obtained in
the simulations. The diagram displays a diagonal cut through the unit cell from one corner to the
opposite corner at two representative bifurcation parameters, $\varepsilon=0$ and $\varepsilon=0.35$. There
is no adjustable parameter in this comparison, apart from centering the peak and normalizing to
the wavelength.
While the theoretical and experimental results differ near the critical value because of the imperfection, as
discussed in section~\ref{sec:scaling}, the data show a perfect match at higher amplitudes~($\varepsilon=0.35$). 

\begin{figure}
\centering
\parbox{0.34\columnwidth}{%
\centering
\includegraphics[width=\linewidth]{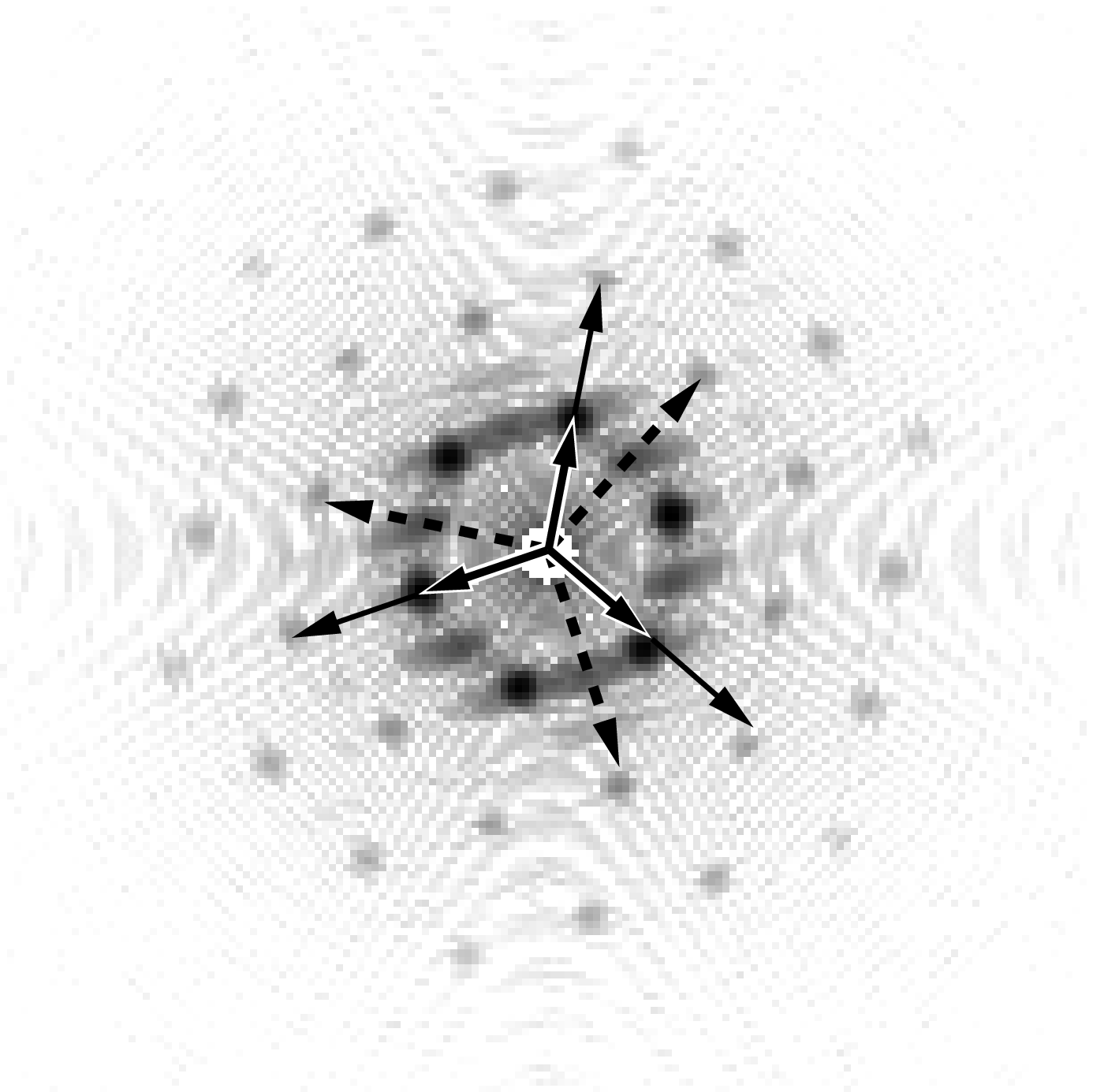}\\
(a)}\hfill
\parbox{0.65\columnwidth}{%
\centering
\includegraphics[width=\linewidth]{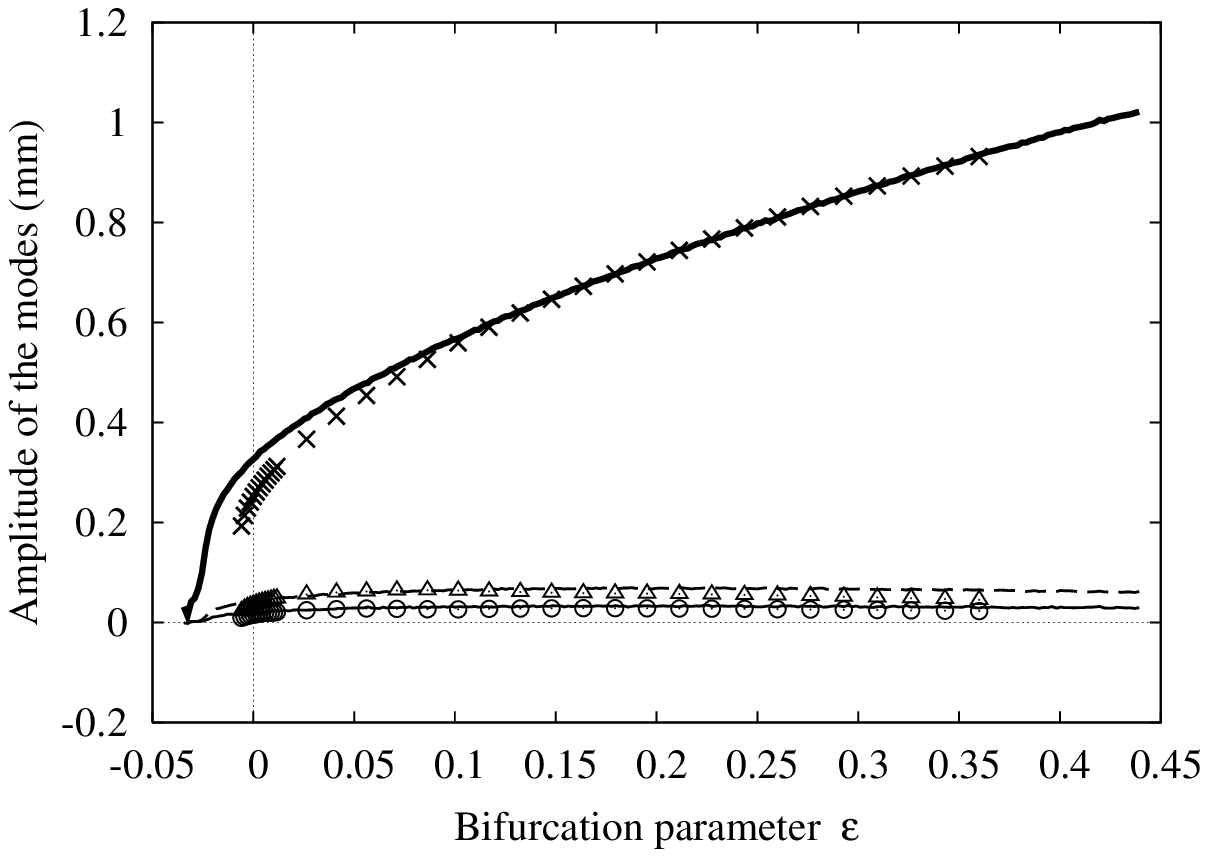}\\
(b)}
\caption{Analyzing the first three harmonics of the pattern. 
Figure~(a) displays a logarithmic greyscale image of the Fourier transform at $B=20.1068\,\mathrm{mT}$.
The wavevectors of the extracted harmonics are denoted by arrows. 
Figure~(b) shows the corresponding amplitudes of the modes. Lines~(symbols) mark experimental~(computed) data, respectively.
Solid thick
lines and crosses represent 
basic modes, solid thin lines and circles modes of the
form $2\vect{k_1}$, dashed lines and triangles modes of the form $\vect{k_1}-\vect{k_2}$. 
}
\label{fig:modes}
\end{figure}
For small amplitudes, it is expected that 
the pattern can be approximated by the dominating Fourier mode, i.e.\ a hexagonal pattern made up of three 
cosine waves:
\[ 
A(\vect x)= \frac{2}{9} \left(\cos \vect k_1 \vect x + \cos \vect k_2 \vect x + 
\cos \vect k_3 \vect x\right)+ \frac{1}{3}.
\]
Here the wavevectors $\vect k_{1,2,3}$ enclose an angle of $120^{\circ}$, sum up to
zero and have the same magnitude $k$. 
To give a quantitative measure to what extent the higher modes are important, we fit a hexagonal
grid with the next two higher harmonics to both the experimental as well as the simulation data. 
The wavevectors are $\vect k_1-\vect k_2, \vect k_2-\vect k_3, \vect k_3-\vect k_1$ with the absolute value 
$\frac{k}{2}\sqrt{15}$ and $2\vect k_1, 2\vect k_2, 2\vect k_3$ with the absolute value $2k$. These
vectors are displayed in figure~\ref{fig:modes}~(a).  
Figure~\ref{fig:modes}~(b) shows the dependence of the components on the bifurcation parameter. As expected, the basic
Fourier mode is the largest component which contributes over $90\,\%$ even for the highest observed amplitude.

\begin{figure}
\parbox{0.49\columnwidth}{%
\centering
\includegraphics[width=\linewidth]{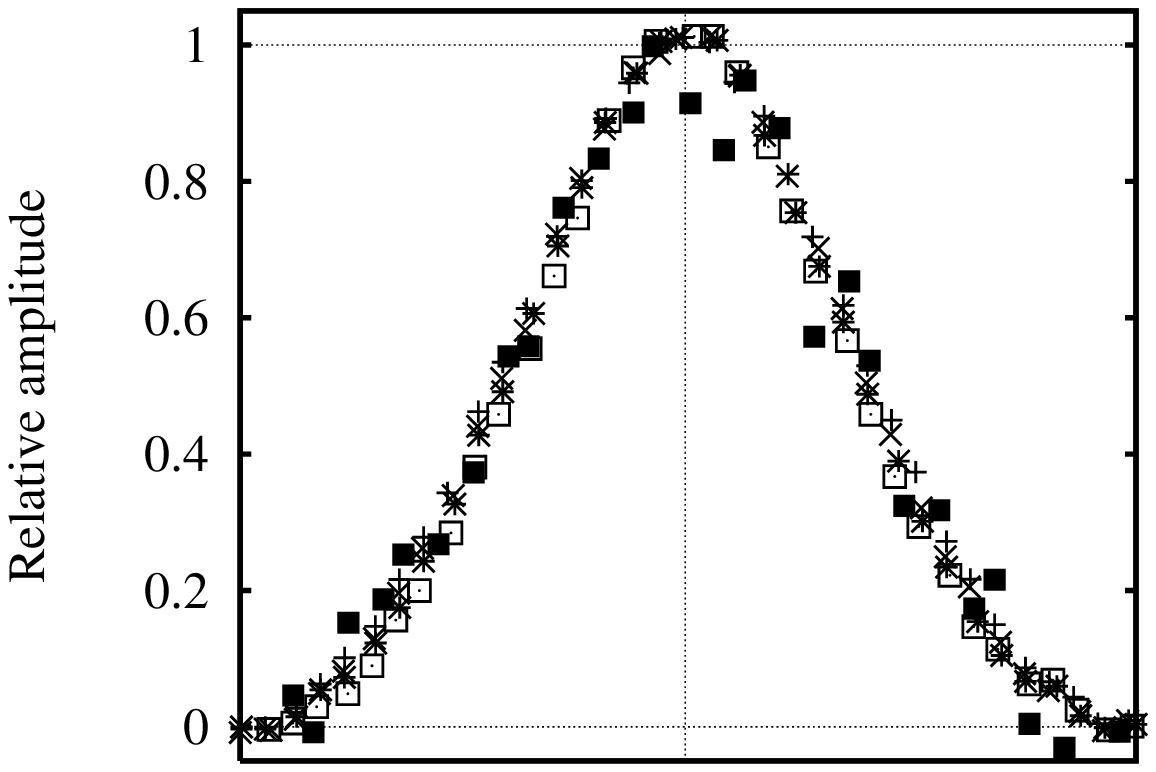}\\
(a)}\hfill
\parbox{0.49\columnwidth}{%
\centering
\includegraphics[width=\linewidth]{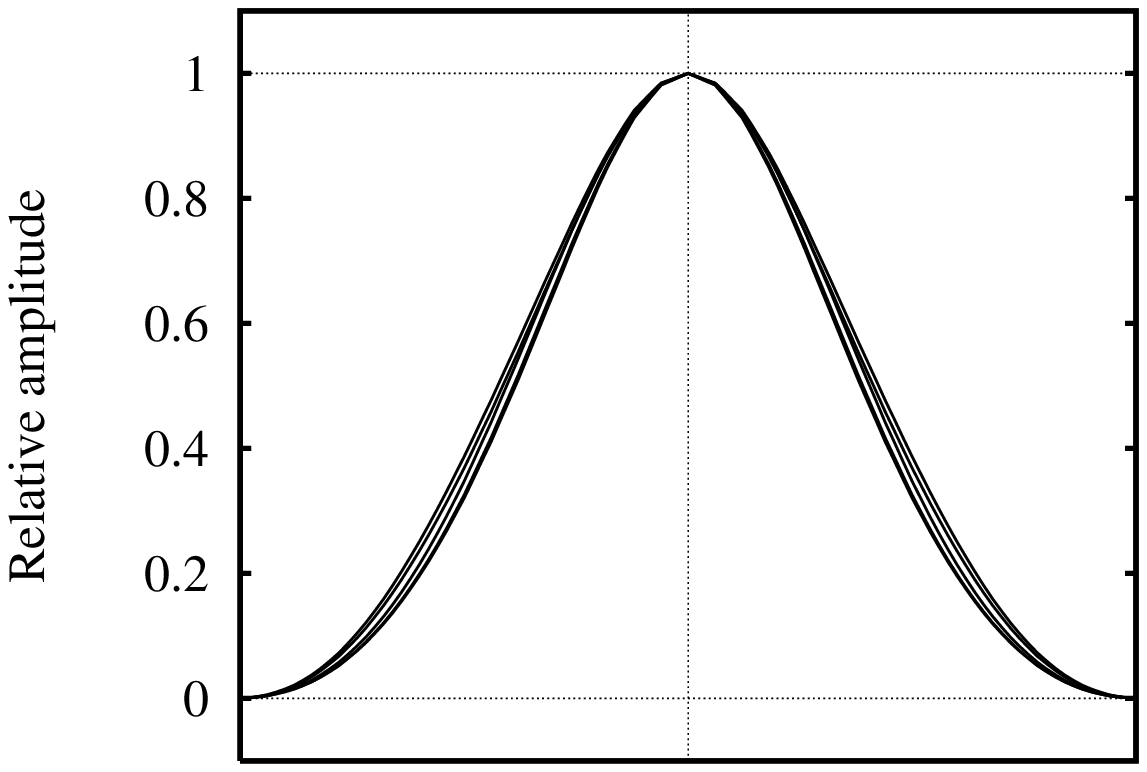}\\
(b)}
\caption{The experimental (a) and numerically obtained (b) normalized profiles of different peaks show that the shape changes only slightly.}
\label{fig:invariantshape}
\end{figure}

Since there is only a small amount of the higher harmonics, the shape of the peaks is mostly
determined by the basic mode and thus should not vary much over the measured range. This can be
seen directly if we rescale both the experimental and numerical data in a way that all peaks have the
same width and height. In figure~\ref{fig:invariantshape}, we plot five different normalized peaks from
the whole range. While the experimental values seem to match perfectly due to the noise, the
numerical data exhibit a certain tendency to sharper peaks for an increasing field. However, this
effect is very small, so the assumption of an invariant shape is a good approximation within the
range $\varepsilon<0.44$. 

\section{Discussion and Conclusion}
We have studied experimentally and numerically the normal field instability in a ferrofluid. 
The qualitative features of this bifurcation can be described very well by 
the nonlinear theory of \cite{friedrichs2001}. A quantitative comparison with this theory is not 
possible, because their approximations are only valid up to
$\chi_0\approx 1$, which is exceeded by the initial susceptibility of our fluid $\chi_0=1.17$.

For a quantitative comparison we had to calculate the surface topography numerically by a finite element method. 
Our computations agreed with the measured amplitude to within $1\,\%$, provided that the uncertainties
of the material parameters and geometrical imperfections were taken into account by matching the critical induction $B_c$.
This indicates that a fluid as complex as magnetic liquids can be indeed described well
by the set of three basic equations, the Navier--Stokes equation, the Maxwell equation and the
Young--Laplace equation.
It turned out to be essential to include the experimentally obtained magnetization curve.

In an attempt to measure the preferred wavenumber -- namely the one minimizing
the free energy -- in the nonlinear regime, we have introduced somewhat
softened boundary conditions in the form of a ramp, which allows for smooth variation 
of the wavenumber as a function of the magnetic field. 
It turned out that this
ramp stabilizes wavenumbers above the critical value of $k_c$, while the theory
by \cite{friedrichs2001} predicts the preferred number to be below $k_c$.
Whether a ramp can be constructed which selects the preferred wavenumber,
remains to be investigated in the future.

Because the radioscopic measurement technique allows us to investigate the full
profile of the peaks, we are able to quantify the ratio of the fundamental mode
to the higher harmonics under variation of $B$. It turned out, that the higher
harmonics contribute less than $10\,\%$ in a range up to $\varepsilon=0.44$.
This is an encouraging result for further analytical treatment of the problem
in terms of amplitude equations for the first few modes.  
Whether the contribution of higher harmonics remains small for fluids with higher 
susceptibility, where already visual inspection reveals sharper peaks, remains a topic 
of further investigations. 

%
%

\begin{acknowledgements}
We thank Ren\'e Friedrichs and Adrian Lange for helpful discussions.
We are grateful to Robert Krau\ss\ and Carola Lepski for measuring the material parameters of the
magnetic fluid, and Klaus Oetter for his construction work. Financial support by Deutsche
Forschungsgemeinschaft grant Ri~1045/1-4
and To~143/4 is gratefully acknowledged. 
\end{acknowledgements}
\bibliographystyle{jfm} 
\bibliography{mfrr,mfcom,comment,xr,numerik}

\end{document}